\documentclass[prd,nofootinbib,11pt,longbibliography]{revtex4}
\usepackage{amssymb,amsmath,latexsym,braket,fancyref,hyphenat,slashed,enumitem,graphicx,mathtools}

\usepackage[caption=false]{subfig}

\newcommand{\mD}{\mathbf{m}_D}
\newcommand{\MD}{\mathbf{M}_D}
\newcommand{\MN}{\mathbf{M}_N}
\newcommand{\m}{\mathbf{m}}
\newcommand{\M}{\mathbf{M}}
\newcommand{\mM}{\mathbf{m}_M}
\newcommand{\Y}{\mathbf{Y}}
\newcommand{\Bnu}{\mathbf{B}_\nu}
\newcommand{\bnu}{\mathbf{b}_\nu}
\newcommand{\Anu}{\mathbf{A}_\nu}

\newcommand{\SL}{{S_L}}

\newcommand{\oSLc}{{\overline{S}^C_L}}
\newcommand{\sSL}{{\tilde{S}_L}}

\newcommand{\snuR}{{\tilde{\nu}_R}}
\newcommand{\onuR}{{\overline{\nu}_R}}
\newcommand{\nuRc}{{\nu^C_R}}

\newcommand{\nuL}{{\nu_L}}
\newcommand{\snuL}{{\tilde{\nu}_L}}

\newcommand{\onuLc}{{\overline{\nu}^C_L}}

\newcommand{\Id}{\mathbf{1}}
\newcommand{\msqL}{m^2_{\tilde{\ell}}}
\newcommand{\msqe}{m^2_{\tilde{e}}}
\newcommand{\msqnu}{m^2_{\tilde{\nu}}}
\newcommand{\wh}{\widehat}
\newcommand{\Hut}{\mathbf{\tilde{H}_1}}
\newcommand{\Hdt}{\mathbf{\tilde{H}_2}}
\newcommand{\nn}{\nonumber}
\newcommand{\mI}{m_{\chi^0_1}}
\newcommand{\mII}{m_{\chi^0_2}}
\newcommand{\mIII}{m_{\chi^0_3}}
\newcommand{\mIV}{m_{\chi^0_4}}

\numberwithin{equation}{section}

\begin{document}

${}$\vspace{-3cm}
\begin{flushright}
MAN/HEP/2020/07\\
August 2020
\end{flushright}

\title{{\LARGE Radiative Neutrino Masses in the {\boldmath $\nu_R$}MSSM}\\[5mm] }

\author {\large Pablo Candia da Silva\footnote{E-mail address: {\tt pablo.candiadasilva@postgrad.manchester.ac.uk}}}

\author {\large Apostolos Pilaftsis\footnote{E-mail address: {\tt apostolos.pilaftsis@manchester.ac.uk}}\\}

\affiliation{\vspace{0.2cm} 
Consortium for Fundamental Physics, School
  of Physics and Astronomy, University of Manchester, Manchester, M13
  9PL, United Kingdom}

\begin{abstract}
\vspace{2mm}
\centerline{\small {\bf ABSTRACT}}
\medskip
\noindent
We present a complete analysis of scenarios with radiatively generated
neutrino masses that may occur in the Minimal Supersymmetric Standard
Model with low-scale right-handed neutrinos. For brevity, we call such
a model the $\nu_R$MSSM.  We pay particular attention to the impact of
the non-renormalization theorem of supersymmetry (SUSY) on the
loop-induced neutrino masses, by performing our computations in the
weak and flavour bases. In~particular, we find that the smallness of
the observed light neutrino masses may naturally arise due to a soft
SUSY-screening effect from a nearly supersymmetric singlet neutrino
sector. The profound phenomenological and cosmological implications
that may originate from this screening phenomenon in the $\nu_R$MSSM and its
minimal extensions are discussed.

\end{abstract}

\maketitle

\section{Introduction}\label{sec:Intro}

Supersymmetry (SUSY)~\cite{Volkov:1973ix,Wess:1974tw} is an elegant
theoretical framework which aspires to technically address several
problems that are central in Particle Physics and Cosmology, such as
the infamous gauge-hierarchy problem, the unification of gauge
couplings and the nature of the Dark Matter~(DM) in the
Universe${}$~\cite{Nilles:1983ge}. In its exact realisation, SUSY is
endowed with powerful non-renormalization theorems that forbid the
presence of new operators in the so-called super\-potential to all
orders in loop expansion~\cite{Grisaru:1979wc}. Even beyond the
tree level, the existing superpotential operators do not require
renormalization other than the one that arises from the wavefunctions
of the fields involved${}$~\cite{Wess:1992cp}. However, for
phenomenological reasons, SUSY needs to be broken at scales larger
than the electroweak (EW) scale, at least higher than
TeV~\cite{Djouadi:2005gj}. If this breaking is `soft' through
holomorphic operators of energy dimensions~3 and less, then this
softly broken SUSY will still help to eliminate all quadratically
sensitive ultra-violet (UV) divergences in the superpotential. This
property is crucial in stabilising the EW scale against quantum
corrections of new physics (other than gravity) that may take place at
much higher scale, e.g.~at scales of gauge-coupling unification in
Grand Unified Theories~(GUTs)~\cite{Dimopoulos:1981zb} 
whose low-energy limit includes the Standard Model~(SM).

A minimal realisation of softly broken SUSY is the so-called Minimal
Supersymmetric Standard Model (MSSM)~\cite{HABER198575}. As a remnant
of the non-renormalization SUSY theorems mentioned above, there exist
now regions of parameter space for which physical observables vanish
or become very suppressed, while they are forbidden in the exact SUSY
limit. Typical examples in which this soft SUSY-screening phenomenon
occurs are lepton- and quark-flavour-violating decay processes, such
as $b\to s\gamma$~\cite{Bertolini:1990if} and
$\mu \to
e\gamma$~\cite{Deppisch:2004fa,Ilakovac:2009jf,Ilakovac:2012sh},
as well as flavour-conserving observables, such as electric and
anomalous magnetic dipole moments of
leptons~\cite{Carena:1996qa,Ilakovac:2013wfa}.

In this paper we show that a similar soft SUSY-screening phenomenon
may be the origin of the smallness of the observed light neutrino
masses~\cite{Fukuda:1998mi,Ahmad:2001an,Ahmad:2002jz} in theories, in
which light neutrino masses are forbidden at the tree level. To
explicitly demonstrate this screening phenomenon, we study typical
scenarios that may occur in the MSSM with a number $n_R$ of
electroweak- or TeV-scale right-handed (singlet) neutrinos $\nu_{iR}$
(with $i = 1,2,\dots,n_R$). For brevity, we call such a model
the~$\nu_R$MSSM. To be able to have good control of the
SUSY-screening effect on the loop-induced neutrino masses, we perform
our computations of the contributing Feynman graphs in the weak
and flavour bases, rather than in the mass basis.

A widely explored framework accounting for the origin of the very
small neutrino masses is given by the so-called seesaw mechanism
\cite{Minkowski:1977sc,Yanagida:1979as,GellMann:1980vs,
  Mohapatra:1979ia,Mohapatra:1980yp,Schechter:1980gr,Lazarides:1980nt,Magg:1980ut,Foot1989}.
This mechanism relies upon the hypothesis that neutrinos are Majorana
fermions~\cite{Majorana:107755}. In its most popular implementation,
the Type-I seesaw
scenario${}$~\cite{Minkowski:1977sc,Yanagida:1979as,GellMann:1980vs,
Mohapatra:1979ia, Mohapatra:1980yp,Schechter:1980gr} postulates the existence of 
right-handed neutrinos, $\nu_{iR}$, which are singlets under the SM
gauge group. This scenario leads to an effective neutrino mass matrix
$\mathbf{m}_\nu$, which is parametrically suppressed by the
lepton-number-violating (LNV) mass matrix of the singlet
neutrinos~$\mM$, i.e.
\begin{equation}
   \label{eq:seesaw}
\mathbf{m}_\nu\ =\ -\mD \mM^{-1} \mD^{\sf T}\,,
\end{equation}
where $\mD$ is the Dirac mass matrix which is generated after
spontaneous symmetry breaking (SSB) by the vacuum expectation value (VEV) of
the SM Higgs doublet $v_{\rm SM} \approx 246$~GeV.  For
$\mD \sim v_{\rm SM}$, the seesaw formula~\eqref{eq:seesaw} implies
that $\mM$ should be of order  $10^{14}$~GeV, namely close to the GUT scale, in
order to account for the observed sub-eV neutrino masses. However, a
phenomenological difficulty of such a scenario is that the required
high-scale of the seesaw mechanism renders its LNV singlet sector not
directly testable in any foreseeable experiment.

The above difficulty may be circumvented in low-scale seesaw
models~\cite{Deppisch:2015qwa}, 
where the LNV scale as dictated by the size of the mass matrix $\mM$
is closer to the EW scale. Typical examples are the inverse seesaw scenario
(ISS)~\cite{Mohapatra:1986aw,Mohapatra:1986bd}, or 
radiative seesaw scenarios where the neutrino masses are absent at the
tree level, but they are generated at loop
level~\cite{Pilaftsis:1991ug,Kersten:2007vk,Dev:2012sg,Grimus:2018rte}.
For a comprehensive review, the interested reader may
consult~\cite{Cai:2017jrq}.  As mentioned above, SUSY-screening is
another mechanism to naturally predict light neutrino masses, without
introducing an unnecessary disparity between the LNV and EW scales,
thereby giving rise to a LNV sector that could be directly probed at
high-energy colliders~\cite{Datta:1993nm,Bray:2007ru,Atre:2009rg,
Cvetic:2010rw,Dev:2013wba,Deppisch:2015qwa,Das:2017gke,Bhardwaj:2018lma}. 
Here, we explicitly demonstrate this phenomenon
within the context of a few representative scenarios in
the~$\nu_R$MSSM.  Unlike previous
studies~\cite{Dedes:2007ef,Hirsch:2009ra,Hollik:2015lwa}, we calculate
the complete set of diagrammatic contributions that come from both the
ordinary SM sector~\cite{Pilaftsis:1991ug} (called here the
$\nu_R$SM) and its SUSY counterpart. In so doing, we pay particular
attention to the impact of the SUSY non-renormalization theorem on the
loop-induced neutrino masses.

The paper is organised as follows. After this introductory section, we
describe the $\nu_R$MSSM in Section~\ref{sec:nuRMSSM}, including our
conventions and notations for the field content. In
Section~\ref{sec:1loop}, we calculate the one-loop induced neutrino
masses, by performing our computations in the weak and flavour
bases. In particular, we check the vanishing of the neutrino masses in
the exact SUSY limit, independently of whether LNV mass parameters are
present in the superpotential or not.  In this way, we can identify
the regions of parameter space for soft SUSY screening. The latter guides our
analysis in Section~\ref{sec:Results}, where we show numerical results
for particular radiative neutrino scenarios in the $\nu_R$MSSM.
Section~\ref{sec:Conclusions} summarises our conclusions, as well as
presents future research directions.  All technical details pertinent
to the $\nu_R$MSSM and our calculations are given in
Appendices~\ref{AppendixHiggsSector}, \ref{AppendixPropagators}
and~\ref{AppendixCoefficients}.

\section{The {\boldmath $\nu_R$} Minimal Supersymmetric Standard
  Model}\label{sec:nuRMSSM} 

As mentioned in the Introduction, the $\nu_R$MSSM is obtained by
adding a number of $n_R$ left-chiral superfields $\wh{N}^C_i$ (with
$i = 1,2,...,n_R$) to the field content of the MSSM. There are several
studies of the $\nu_R$MSSM in the
literature~\cite{Hisano:1995nq,Grossman:1997is,Casas:2001sr,Farzan:2003gn,Chun:2005qw,Demir:2005ya,Arganda:2005ji,Dedes:2007ef,Heinemeyer:2010eg}. In
our paper, we adhere to the notation displayed in
Table~\ref{Tab:nuR_MSSM_content} for the full $\nu_R$MSSM spectrum of
fields.

\begin{table}[t]
	\begin{tabular}{|c||c|c|c|}
		\hline
		Superfields & Bosons & Fermions & SU(3)$_c$$\otimes$SU(2)$_L$$\otimes$U(1)$_Y$\\\hline\hline
		\underline{Gauge multiplets}& & &\\[0.3ex]
		$\wh{G}^a$ & $G^a_\mu\frac{1}{2}\lambda^a$ & $\tilde{g}^a$ & $(\textbf{8},\textbf{1},0)$\\[0.3ex]
		$\wh{W}^a$ & $W^i_\mu\frac{1}{2}\sigma_i$ & $\tilde{W}^i$ & $(\textbf{1},\textbf{3},0)$\\[0.3ex]
		$\wh{B}$ & $B_\mu$ & $\tilde{B}$ & $(\textbf{1},\textbf{1},0)$\\
		&&&\\\hline
		\underline{Matter multiplets}& & &\\[0.3ex]
		$\wh{L}$ & $\tilde{L}^{\sf T} = (\snuL,\tilde{e}_L)$ & $L^{\sf T} = (\nuL,e_L)$ & $(\textbf{1},\textbf{2},-1)$\\[0.3ex]
		$\wh{E}^C$ & $\tilde{e}^*_R$ & $e^C_R$ & $(\textbf{1},\textbf{1},2)$\\[0.3ex]
		$\wh{Q}$ & $\tilde{Q}^{\sf T} = (\tilde{u}_L,\tilde{d}_L)$ & $Q^{\sf T} = (u_L,d_L)$ & $(\textbf{3},\textbf{2},1/3)$\\[0.3ex]
		$\wh{U}^C$ & $\tilde{u}^*_R$ & $u^C_R$ & $(\textbf{3},\textbf{1},-4/3)$\\[0.3ex]
		$\wh{D}^C$ & $\tilde{d}^*_R$ & $d^C_R$ & $(\textbf{3},\textbf{1},2/3)$\\[0.3ex]
		$\wh{H}_d$ & $H^{\sf T}_d=(H^0_d,H^-_d)$ & $\tilde{h}^{\sf T}_d=(\tilde{h}^0_d,\tilde{h}^-_d)$ & $(\textbf{1},\textbf{2},-1)$\\[0.3ex]
		$\wh{H}_u$ & $H^{\sf T}_u=(H^+_u,H^0_u)$ & $\tilde{h}^{\sf T}_u=(\tilde{h}^+_u,\tilde{h}^0_u)$ & $(\textbf{1},\textbf{2},1)$\\[0.3ex]
		$\wh{N}^C$ & $\snuR^*$ & $\nu^C_R$ & $(\textbf{1},\textbf{1},0)$\\
		&&&\\\hline
	\end{tabular}
	\caption{Particle content of the $\nu_R$MSSM. The numbers in
          boldface indicate the dimension of the gauge group
          representation under which each multiplet transforms. Here,
          $\sigma_i$ with $i=1,2,3$ are the usual Pauli matrices,
          while $\lambda^a$ with $a=1,2,...,8$ are the Gell-Mann
          matrices.} 
	\label{Tab:nuR_MSSM_content}
\end{table}

The superpotential of the $\nu_R$MSSM is given by 
\begin{equation}
  \label{superpotential}
W\ =\ W_{\text{MSSM}} + \wh{L}i\sigma_2\wh{H}_u \Y_\nu \wh{N}^C + 
\frac{1}{2}\wh{N}^C \mM \wh{N}^C\,,
\end{equation}
where $W_{\text{MSSM}}$ is the usual MSSM superpotential, $\Y_\nu$ and
$\mM$ denote the $3\times n_R$ neutrino Yukawa matrix and the
$n_R\times n_R$ Majorana mass matrix, respectively.  These latter
matrices also appear in the $\nu_R$SM. Note that we use boldface format
to highlight matrices with flavour structure. Employing\- 
superspace techniques~\cite{1978ForPh..26...57S}, we can 
derive the SUSY Yukawa Lagrangian of interest to us,
\begin{align}
    \label{YukMSSMnuR}
  -\mathcal{L}_{\rm Y} \ =\ \overline{L}^Ci\sigma_2 H_u \Y_\nu \nuRc +
  \tilde{L}^{\sf T}i\sigma_2 \tilde{h}_u \Y_\nu \nuRc + \overline{L}^C
  i\sigma_2 \tilde{h}_u \Y_\nu \snuR^* + \frac{1}{2}\onuR \mM \nuRc +
  \textrm{H.c.} 
\end{align}
Likewise, from the superpotential~\eqref{superpotential}, one may
derive the $F$-term contributions from the right-handed
sneutrinos~$\snuR$ to the scalar potential, 
\begin{align}
    \label{SPotMSSMnuR}
V^{\snuR}_F\ =\ &\, \snuR^* \Y_\nu^{\sf T} \Y_\nu^*\snuR H^\dagger_u H_u -
               \Big(\tilde{e}^*_R \Y_e^{\sf T} \Y_\nu^*\snuR H^\dagger_u H_d
               + \textrm{h.c.}\Big) + \tilde{L}^{\sf T} \Y_\nu\snuR^*
               \,\snuR \Y_\nu^\dagger \tilde{L}^* \nonumber\\ 
&- \left(\mu\snuR \Y_\nu^\dagger\tilde{L}^\dagger H_d +
  \textrm{h.c.}\right) - H^{\sf T}_ui\sigma_2 \tilde{L}\Y_\nu \Y_\nu^\dagger
  \tilde{L}^\dagger i\sigma_2 H^*_u + \left(\tilde{L}^{\sf T} i\sigma_2 H_u
  \Y_\nu \mM^\dagger\snuR + \textrm{H.c.}\right)\nonumber\\ 
&+ \snuR^* \mM \mM^\dagger \snuR\, . 
\end{align}
Observe that the $F$-term induced potential~$V^{\snuR}_F$ contains the
LNV operator, $\tilde{L}^{\sf T} i\sigma_2 H_u\snuR$ given by the
penultimate term on the RHS of~\eqref{SPotMSSMnuR}, which violates the
lepton number by two units. As~we~will discuss in the next section,
this SUSY-generated LNV operator plays an instrumental role in the
determination of the radiative neutrino masses by screening the effect
of the ordinary right-handed neutrinos from the $\nu_R$SM.

Finally, the relevant soft SUSY-breaking Lagrangian derivable from the
superpotential~\eqref{superpotential} is given by
\begin{equation}
-\mathcal{L}_{\rm soft} =\ 
-\mathcal{L}^{\rm MSSM}_{\rm soft} + \snuR^* \m^2_{\tilde{\nu}}\snuR +
\Big(\snuR^* \,\bnu \mM\snuR^* + \tilde{L}^{\sf T}i\sigma_2 H_u \Y_\nu\Anu
\snuR^* + \textrm{H.c.}\Big), 
\end{equation}
where $\mathcal{L}^{\rm MSSM}_{\rm soft}$ denotes the usual soft
SUSY-breaking contribution from the MSSM, and $\m^2_{\tilde{\nu}}$,
$\bnu$ and $\Anu$ are $n_R\times n_R$-dimensional matrices.

A minimal radiative seesaw scenario of the Type I is the ISS
realisation presented in~\cite{Dev:2012sg}.  As in the standard ISS
model~\cite{Mohapatra:1986aw,Mohapatra:1986bd}, the radiative model
has new fermionic singlets that come in $n$ pairs, i.e.~$n_R = 2n$.
Thus, the model contains $n$ right-handed neutrinos $\nu_{iR}$, with
$i = 1,2,\dots, n$, while the remaining $n$ right-handed neutrinos are
renamed as $\nu_{i R} \equiv (S_{\alpha L})^C$, with
$i = n +\alpha = n+1,n+2,\dots, 2n$.  In terms of the fields
$\big(\nu_{1,2,3L},\, \nu^C_{iR},\, S_{\alpha L}\big)$, with
$i,\alpha = 1,2,\dots, n_R$, the tree-level neutrino-mass Lagrangian
after SSB reads \renewcommand\arraystretch{1.2}
\begin{align}
    \label{generalISS}
-\mathcal{L}^\nu_{\rm mass}
=& \frac{1}{2}\Big(\onuLc,\onuR,\oSLc\Big)\begin{pmatrix}
\mathbf{0}_3 & \MD & \mathbf{0}_{3\times n}\\
\MD^{\sf T} & \boldsymbol{\mu}_R & \MN^{\sf T}\\
\mathbf{0}_{n\times 3} & \MN & \boldsymbol{\mu}_S
\end{pmatrix}
\begin{pmatrix}
\nuL\\\nuRc\\\SL
\end{pmatrix}\ +\ \text{H.c.}\,,
\end{align}
with $\MD = \frac{v_u}{\sqrt{2}}\Y_\nu$. The standard seesaw matrix of
the $\nu_R$SM is recovered if the following identifications are made:
\begin{align}
   \label{eq:mM-ISS}
\mD\ \equiv\ \Big(\MD \,,\, \mathbf{0}_{3\times n}\Big)\,,\qquad
   \mM \equiv \begin{pmatrix}
\boldsymbol{\mu}_R & \MN^{\sf T}\\
\MN & \boldsymbol{\mu}_S
\end{pmatrix}.
\end{align}
The mass matrix of~\eqref{generalISS} has two {\em soft} LNV matrix-valued
parameters, $\boldsymbol{\mu}_{R,S}$. In the limit where lepton number
is preserved $\boldsymbol{\mu}_{R,S} = {\bf 0}_n$, the
neutrino mass eigenstates become exact Dirac states to all loop
orders. The usual ISS scenario, which has been studied widely in the
literature~\cite{Banerjee:2013fga,Hirsch:2009ra,An:2011uq,Guo:2013sna,
  Arganda:2014dta}, is obtained for
$\boldsymbol{\mu}_{S} \neq {\bf 0}_n$ and
$\boldsymbol{\mu}_R = {\bf 0}_n$. However, our interest here
is the SUSY extension of the {\em radiative} ISS model
in~\cite{Dev:2012sg}, for which
$\boldsymbol{\mu}_S = {\bf 0}_n$ and
$\boldsymbol{\mu}_R \neq {\bf 0}_n$.

In the context of the $\nu_R$MSSM, left and right-handed sneutrinos
mix. In the weak basis $(\snuL,\snuR^*,\snuL^*,\snuR)^{\sf T}$, 
the following sneutrino mass matrix may be derived:
\begin{align}
   \label{SleptonMassMatrix}
    \mathbf{M}^2_{\tilde{\nu}} = \begin{pmatrix}
    \mathbf{H}_1 & \mathbf{N} & \mathbf{0} & \mathbf{M}\\
    \mathbf{N}^\dagger & \mathbf{H}^{\sf T}_2 & \mathbf{M}^{\sf T} & \mathbf{B}^\dagger_\nu\\
    \mathbf{0} & \mathbf{M}^* & \mathbf{H}^{\sf T}_1 & \mathbf{N}^*\\
    \mathbf{M}^\dagger & \mathbf{B}_\nu & \mathbf{N}^{\sf T} & \mathbf{H}_2
    \end{pmatrix},
\end{align}
where
\begin{subequations}
\begin{align}
    \mathbf{H}_1 =&\,\mathbf{m}^2_{\tilde{\ell}} + \mD^* \mD^{\sf T} +
                    \frac{1}{2}M^2_Z\cos 2\beta\Id_3\,,\label{H1}\\ 
    \mathbf{H}_2 =&\, \mathbf{m}^2_{\tilde{\nu}} + \mD^{\sf T} \mD^* + \mM
                    \mM^\dagger\label{H2}\,,\\[1ex] 
    \mathbf{M} =&\, \mD^*(\Anu^* - \mu \cot{\beta}\,\Id_{n_R})\,,\label{Minsertion}\\[1ex]
    \mathbf{N} =&\, \mD^* \mM\,,\label{Ninsertion}\\[1ex]
    \mathbf{B}_\nu =&\, \bnu \mM\,.
\end{align}
\end{subequations}
Note that the sneutrino mass matrix for the ISS scenario, in
which the fields $\sSL$ are included in conjunction with~$\snuR^*$, can
straightforwardly be obtained after making the identifications for the
mass matrices~$\mM$ and $\mD$ mentioned above [cf.~\eqref{eq:mM-ISS}].

\section{Radiative neutrino masses}\label{sec:1loop}

In this section, we calculate the relevant Feynman-diagrammatic
contributions to the neutrino self-energy matrix,
$\Sigma({\slashed{p}})$. The calculation is carried out in the flavour
and weak bases, rather in the mass basis, in order to avoid the use of
large dimensional matrices describing neutralino and sneutrino mixings.
This enables us to have good control of the soft SUSY-screening
phenomenon mentioned in the Introduction.

\begin{figure}[t]
    \centering
\includegraphics[width=15.cm]{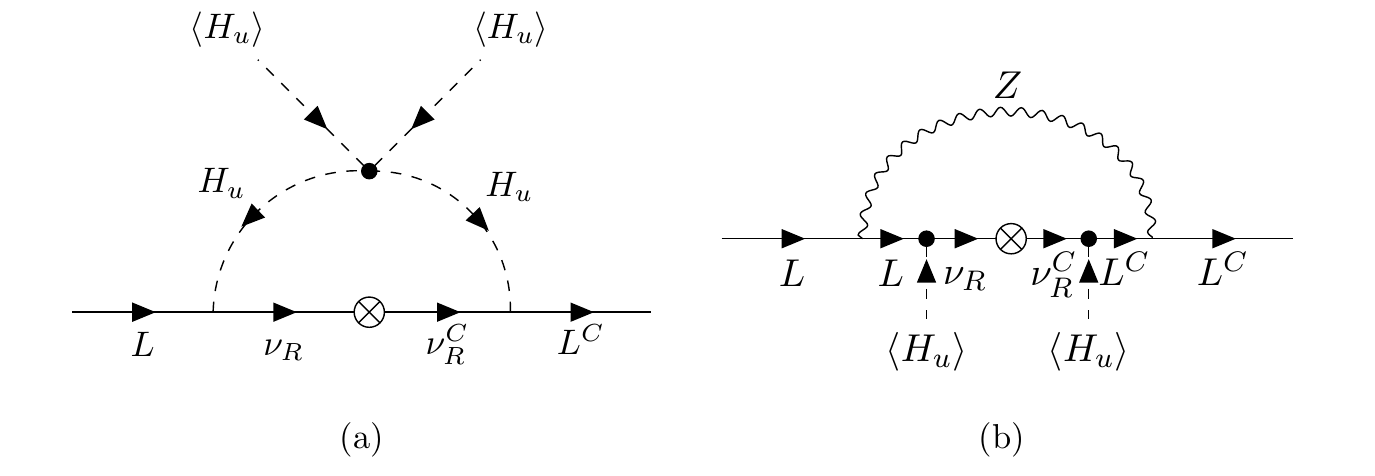}
        \caption{Leading-order diagrams in the $\nu_R$SM that contribute
          to neutrino masses. An encircled cross indicates a LNV
          insertion, a heavy dot in diagram~(a) stands for
          $\lambda_{\rm eff}$ [cf.~\eqref{eq:lambda_eff}], while a dot
          in diagram~(b) represents an $\mathbf{m}_D$ insertion
          [cf.~\eqref{Minsertion}]. Note that the arrows on the
          Higgs-propagator lines indicate hypercharge${}$~flow.}
    \label{SMdiagrams}
\end{figure}

\begin{figure}[t]
\centering
	\includegraphics[width=12.5cm]{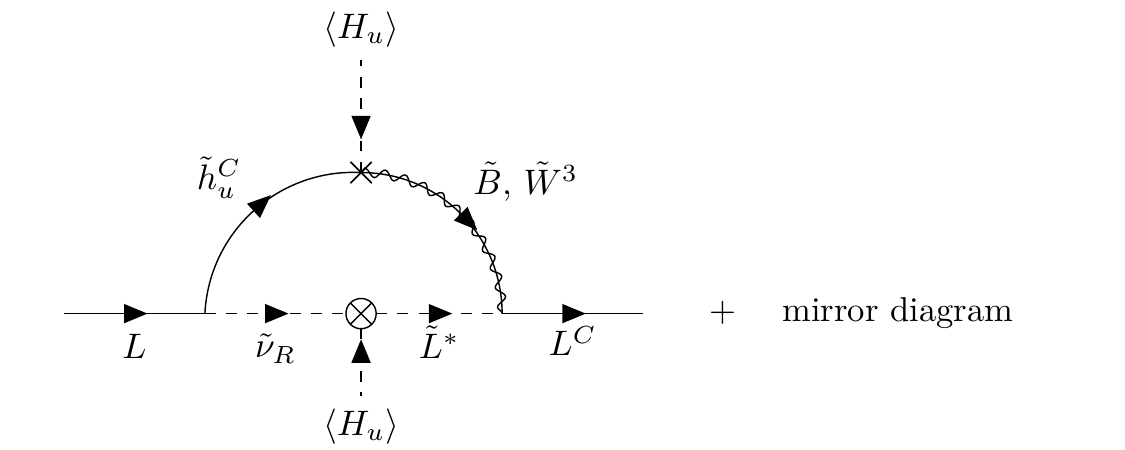}
    \caption{Supersymmetric one-loop contributions to neutrino
      masses. The cross represents a chirality flip in the
      higgsino--gaugino fermionic
      line, and the encircled cross represents
      an $F$-term LNV mass insertion [cf.~\eqref{Ninsertion}].} 
    \label{Supersymmetric graphs}
\end{figure}

\begin{figure}[t]
\includegraphics[width=15cm]{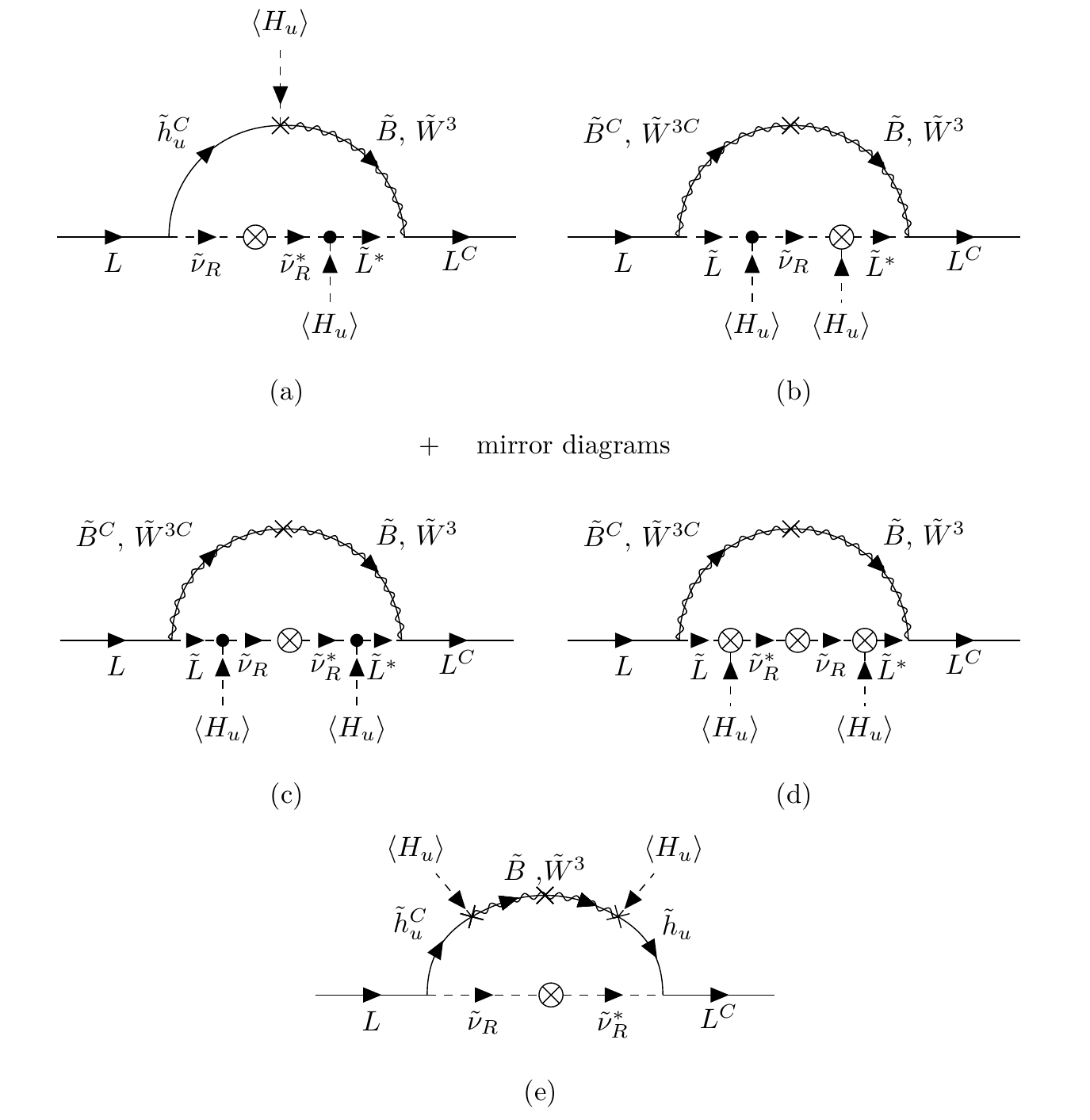}
\caption{Feynman diagrams contributing to neutrino masses when SUSY is
  softly broken. Note that the right-handed sneutrino propagator has a
  LNV mass insertion proportional to $\Bnu$.}
	\label{SoftBreakingDiagrams1}
\end{figure}

For theories with Majorana fermions that we have been considering here, the
neutrino self-energy matrix assumes the general 
form~\cite{Kniehl:1996bd,Pilaftsis:2002nc}: 
\begin{align}
    \Sigma({\slashed{p}})\ =\ \Sigma_L(p^2)\,\slashed{p} P_L\: +\: 
\Sigma_R(p^2)\, \slashed{p} P_R\: + \: \Sigma_M(p^2) P_L\: + \Sigma^*_M(p^2) P_R\, ,
\end{align} 
with $\Sigma_{L,R}(p^2) = \Sigma^\dagger_{L,R}(p^2)$,
$\Sigma_L(p^2) = \Sigma^*_R(p^2)$ and
$\Sigma_M(p^2) = \Sigma^{\sf T}_M(p^2)$. Since
our interest lies only in computing the left-handed effective neutrino
mass matrix $\M_{\nu_L}$, which is generated radiatively, this can
easily be determined by
\begin{align}
    \label{NuMassMatrix}
\M_{\nu_L}P_L\ =\ -P_L\Sigma(\slashed{p}) P_L\Big|_{p\to 0}\ =\  -\Sigma_M(0)P_L\,.
\end{align}
In the weak and flavour spaces, the leading-order contributions to
$\Sigma_M(0)$ are shown in
Figs.~\ref{SMdiagrams}--\ref{SoftBreakingDiagrams1}. Note that
chirality flipping insertions are denoted with a cross on a fermion
line.  If this chirality flip happens to violate the lepton number as
well, we indicate this with a cross inside a circle. In particular, the
analytic results derived from the weak- and flavour-space diagrams
have the advantage that they only depend on the parameters of the
Lagrangian and the particle masses, but {\em not} on mixing matrix elements.

To avoid too large effects of charged lepton flavour violation, we
assume that the soft SUSY-breaking bilinear and trilinear parameters
 are universal at some low-energy scale, i.e.
\begin{align}
   \label{eq:SoftU}
    \m^2_{\tilde{e}} = \msqe \Id_3\,,\qquad \m^2_{\tilde{\ell}} = \msqL
  \Id_3\,, \qquad \m^2_{\tilde{\nu}} = \msqnu \Id_{n_R}\,,\qquad \Anu = A_\nu\Id_{n_R}\,,
\end{align}
so that we can ignore renormalization-group effects for simplicity.
Moreover, we work in second order approximation for the neutrino
Yukawa matrix $\Y_\nu$. This is a good approximation provided the
mixing between light and heavy neutrinos is reasonably small.  To
simplify matters, our computation\- is based on the working
hypothesis that the bilinear mass matrix $\bnu$ is universal,
\begin{equation}
    \label{eq:bnu}
\bnu\ =\ b_\nu\,\Id_{n_R}\, .
\end{equation}
It should be noted here that although the final analytic
expression~$\Sigma_M(0)$ is gauge-fixing parameter independent, the
computation\- of the individual diagrams is done in the Landau gauge.

Let us first consider the Feynman graphs shown in
Fig.~\ref{SMdiagrams} which are closely related to those evaluated in
the $\nu_R$SM~\cite{Pilaftsis:1991ug}. At the one-loop level, there
are two contributions that involve: (a)~the up-type Higgs-boson $H_u$
and (b)~the $Z$ boson. For the diagram in Fig.~\ref{SMdiagrams}(a), we
must notice that it receives significant quantum corrections beyond
the one-loop level, through the operator
$\lambda_{\rm eff} (H^\dagger_uH_u)^2$ whose
coupling~$\lambda_{\rm eff}$ gets considerably
enhanced~\cite{Ellis:1990nz,Haber:1990aw,Okada:1990vk} beyond its SUSY
tree-level value, i.e.~$\lambda_{\rm tree} = (g^2 + g'^2)/8$.  The
effective coupling $\lambda_{\rm eff}$ helps to raise the value of the
lightest CP-even Higgs mass, $m_h$, to that of the observed SM-like
Higgs resonance, i.e.~$m_h \approx
125$~GeV~\cite{Sirunyan:2020xwk}.
For instance, for moderate values of $\tan\beta$,
e.g.~$\tan\beta \stackrel{<}{{}_\sim} 20$, one-loop scalar top
($\tilde{t}$) effects become dominant, leading to an effective
coupling (see, e.g.~\cite{Pilaftsis:1999qt}),
\begin{equation}
    \label{eq:lambda_eff}   
\lambda_{\rm eff}\ =\ \lambda_{\text{tree}}\, \bigg[ 1 -
\frac{3}{8\pi^2} \ln\bigg(
\frac{M^2_{\tilde{t}}}{\overline{m}^2_t}\bigg)\, \bigg]\: +\:
\frac{3y^4_t}{16\pi^2}\, \bigg[
\ln\bigg(\frac{M^2_{\tilde{t}}}{\overline{m}^2_t}\bigg)\:
+\: \frac{|A_t|^2}{M^2_{\tilde{t}}}\, \bigg( 1\, -\,
\frac{|A_t|^2}{12\,M^2_{\tilde{t}}}\bigg)\,\bigg]\, ,
\end{equation}
where $\overline{m}_t$ is the top-quark pole mass, $y_t$ is the
top-quark Yukawa coupling, $M^2_{\tilde{t}}$ is the average of the
soft scalar-top masses squared, and $A_t$ is the respective soft
trilinear coupling. In the present study, we will not specify the full
soft SUSY-breaking sector of the $\nu_R$MSSM, but only adjust
$\lambda_{\rm eff}$, so as to have $m_h = 125.38\pm0.14$~GeV
[cf.~Table~\ref{tab:Benchmark}]. This can always be achieved by an
appropriate choice of the soft parameters associated with the scalar
quark sector of the MSSM~\cite{Lee:2007gn,Heinemeyer:2007aq,Carena:2015uoe}.
Further discussion is given in Appendix~\ref{AppendixHiggsSector}.

Taking the above into account, as well as the pertinent $H^*_uH^*_u$
entry of the Higgs-boson propagator\- matrix (see
Appendix~\ref{AppendixPropagators} for more details), the $H_u$-mediated
graph in Fig.~\ref{SMdiagrams}(a) is found to be
\begin{align}
    \label{SelfHiggs-I} 
i\Sigma^{\text{[1(a)]}}_M\ =\ 2\lambda_{\rm eff}\, \mD
  \mM^\dagger \int\!\frac{d^d k}{(2\pi)^d}\frac{\big(k^2 + m_A^2
  \cos2\beta\big)^2}{k^2(k^2\Id_{n_R} - \mM \mM^\dagger)(k^2 - m^2_A)(k^2
  - m^2_h)(k^2 - m^2_H)}\mD^{\sf T}\,,
\end{align}
where $\lambda_{\rm eff}$ is the effective quartic coupling defined
in~\eqref{eq:lambda_eff}.  

\vfill\eject

Since all loop integrals are evaluated at zero external momentum, it
becomes more convenient to express them in terms of the functions
\begin{align}
   \label{VeltmanIn}
    I_n(m^2_1,m^2_2,\dots,m^2_n) \equiv \int\!\frac{d^d
  k}{(2\pi)^d}\, \prod^n_{j=1}\,\frac{1}{k^2 - m^2_j}\ .
\end{align}
The functions $I_n$ satisfy the recursive relations
\begin{align}
I_n (m^2_1,m^2_2,\dots,m^2_n)\ =\ \frac{\,I_{n-1}(m^2_1,m^2_3,...,m^2_n)\: -\: 
I_{n-1}(m^2_2,m^2_3,\dots,m^2_n)}{\!m^2_1 - m^2_2}\ .\label{VeltmanRecursion}
\end{align}
They are related to the usual Veltman
functions~\cite{Passarino:1978jh} by a multiplicative factor. For
instance, for~$n=2$, we have
\begin{align}
    I_2(m^2_1,m^2_2) = \frac{i}{16\pi^2}\, B_0(0, m^2_1,m^2_2),
\end{align}
where in $d= 4-2\epsilon$ dimensions, 
\begin{align}
   B_0(0,m^2_1,m^2_2)\ =\ C_{\rm UV}\: -\: \frac{m^2_1}{m^2_1 -
  m^2_2}\ln\bigg(\frac{m^2_1}{m^2_2}\bigg)\: -\:
  \ln\bigg(\frac{m^2_2}{\mu^2}\bigg)\: +\: 1\, .\label{I2B0}  
\end{align}
Here, $C_{\rm UV} = \frac{2}{\epsilon} + \ln 4\pi - \gamma_{\rm E}$ is an UV
constant, $\gamma_{\rm E}$
is the Euler--Mascheroni constant, and $\mu$ is an arbitrary mass scale
introduced by 't Hooft.  As we will see below, all loop contributions
yield combinations of functions $I_n$, with $n\ge 3$. In light
of~\eqref{VeltmanRecursion}, this means that the UV divergent parts
depending on $C_{\rm UV}$ cancel out completely, as required by the renormalisability of
the theory.

By virtue of the $I_n$ functions defined in~(\ref{VeltmanIn}), we may 
now re-express~(\ref{SelfHiggs-I}) as follows:
\begin{align}
    i\Sigma^{\text{[1(a)]}}_M =\ &2\lambda_{\rm eff}\,\mD
                           \mM^\dagger\Big[ I_3(m^2_h, m^2_H, \mM
                           \mM^\dagger)\nonumber\\[1.2ex] &+ m^2_A(1 +
                                                            2\cos2\beta)
                                                            I_4(m^2_A,
                                                            m^2_h,
                                                            m^2_H ,\mM
                                                            \mM^\dagger)\nonumber\\[1.5ex] 
    &+ m^4_A \cos^2 2\beta\, I_5(0, m^2_A, m^2_h, m^2_H, \mM \mM^\dagger)\Big]\mD^{\sf T}\,.
    \label{SMhiggsContribution} 
\end{align}
Likewise, the $Z$-boson mediated diagram in Fig.~\ref{SMdiagrams}(b)
is given by
\begin{align}
    i\Sigma^{\text{[1(b)]}}_M\ =\ \frac{3}{4}(g^2 + g'^2)\, \mD
  \mM^\dagger I_3(0, M^2_Z, \mM \mM^\dagger)\,\mD^{\sf
  T}\, .\label{SMzContribution} 
\end{align}

Let us now turn our attention to the respective supersymmetric
contributions to radiative neutrino masses shown in
Fig.~\ref{Supersymmetric graphs}.
The evaluation of these graphs can be done using the 
neutralino propagator matrix in the weak basis, by following the procedure
outlined in Appendix~\ref{AppendixPropagators}.  Thus, the loop integral
for the SUSY graphs in Fig.~\ref{Supersymmetric graphs}  reads
\begin{align}
    i\Sigma^{[2]}_M\ =\
  &\frac{1}{v_u}\int\!\frac{d^dk}{(2\pi)^d}\, \frac{1}{k^2\Id_3
    - \Hut^{\sf T}}\mathbf{N}^*\frac{1}{k^2\Id_{n_R} -\Hdt}\mD^{\sf
    T}\,\frac{g\big(M_{\chi^0}\, \text{adj}\mathcal{F}\big)_{\tilde{h}^0_u
    \tilde{W}^3} -g'\big(M_{\chi^0}\:
    \text{adj}\mathcal{F}\big)_{\tilde{h}^0_u\tilde{B}}}{\det \mathcal{F}}\nn\\[2ex]  
    &\qquad+ \text{ transpose},\label{SusyInt1} 
\end{align}
where $\mathcal{F}$ is a $4\times 4$ matrix related to the weak-space
neutralino propagator and $\text{adj}\mathcal{F}$ stands for its
adjunct. More details may be found in
Appendix~\ref{AppendixPropagators}. In~\eqref{SusyInt1}, we defined
$\tilde{\mathbf{H}}_{{1,2}} \equiv
\mathbf{H}_{{1,2}}|_{\mD=\boldsymbol{0}}$, consistent with our
second-order expansion in the neutrino
Yukawa matrix $\Y_\nu$. Since $\Hut$ is proportional to the
identity in the same order of approximation, we may write 
$\Hut^{\sf T}=\tilde{H}^*_1\Id_3$. Taking the above simplifications into account,
the loop integral may successively be evaluated as follows:
\begin{align}
   \label{SUSYcontribution}
    i\Sigma^{[2]}_M =\ &\frac{1}{v_u}\mathbf{N}^*\int\!\frac{d^d
                          k}{(2\pi)^d}\frac{g(M_{\chi^0}\, \text{adj}
                          \mathcal{F})_{\tilde{h}^0_u \tilde{W}^3} -
                          g'(M_{\chi^0}\, \text{adj}
                          \mathcal{F})_{\tilde{h}^0_u
                          \tilde{B}}}{\det\mathcal{F}\,(k^2 -
                          \tilde{H}_1^*)(k^2\Id_{n_R} - \Hdt)}\mD^{\sf
                          T}\nn 
    \,\,\,+ \text{ transpose}\\[3mm]
    =\ & \mD\mM^\dagger\int\!\frac{d^d k}{(2\pi)^d}\frac{Ak^6 + Bk^4 +
         Ck^2 + D}{\det \mathcal{F}\,(k^2 - \tilde{H}_1^*)(k^2\Id_{n_R} - \Hdt)}\mD^{\sf T}\nn
    \,\,\,+ \text{ transpose}\nonumber\\[3mm] 
 =\ & \mD \mM^\dagger\bigg[ A\, I_3\left(\tilde{H}_1^*,\Hdt, \mIV^2\right)\nonumber
    + \mathcal{B}_{A,B} I_4\left(\tilde{H}_1^*,\Hdt,m^2_{\chi^0_{3}},\mIV^2\right)\nonumber\\[3mm]
    &+ \mathcal{C}_{A,B,C}
      I_5\left(\tilde{H}_1^*,\Hdt,m^2_{\chi^0_{2}},\mIII^2,\mIV^2\right)\nonumber
      + \mathcal{D}_{A,B,C,D} I_6\left(\tilde{H}_1^*,\Hdt,
      \mI^2,\mII^2,\mIII^2,\mIV^2\right)\bigg]\,\mD^{\sf T}\nn\\[2ex] 
    &+ \text{ transpose}, 
\end{align}
with
\begin{align}
   \label{eq:calBCD}
    \mathcal{B}_{A,B}\ =\ & \,A\left(\mI^2 + \mII^2 + \mIII^2\right) + B,\nonumber\\
    \mathcal{C}_{A,B,C}\ =\ & \,A\left(\mI^4 + \mI^2 \mII^2 + \mII^4\right) + B\left(\mI^2 + \mII^2\right) + C,\\
    \mathcal{D}_{A,B,C,D}\ =\ & \,A\,\mI^6 + B\,\mI^4 + C\,\mI^2 + D.\nonumber
\end{align}
Note that the analytic forms of the coefficients $A,B,C$ and $D$ are
given in~\eqref{coefs}.

It can now be verified that the sum of the self-energy expressions
in~\eqref{SMhiggsContribution}, \eqref{SMzContribution} and
\eqref{SUSYcontribution} vanishes in the exact SUSY limit, which is
realised for $\tan\beta = 1$, $\mu=0$ and
$M_{\text{SUSY}}\rightarrow 0$, with~$\lambda_{\rm eff} \to \lambda_{\rm
  tree}$, according to our discussion
in~Appendix~\ref{AppendixHiggsSector}.

Finally, we must take into consideration weak- and flavour-space
graphs, as depicted in Fig.~\ref{SoftBreakingDiagrams1}, which are generated
when SUSY is softly broken. We start by calculating the loop integral
for the diagram in Fig.~\ref{SoftBreakingDiagrams1}(a). This is given by
\begin{align}
    \label{SoftInt1}
i\Sigma^{\text{[3(a)]}}_M =\ &\frac{1}{v_u}\int\!\frac{d^d
                                k}{(2\pi)^d}\, \frac{1}{k^2\Id_3
                                - \Hut^{\sf
                                T}}\mathbf{M}^*\frac{1}{k^2\Id_{n_R} -
                                \Hdt^{\sf
                                T}}b_\nu^*\mM^\dagger\frac{1}{k^2\Id_{n_R}
                                - \Hdt}\mD^{\sf T}\nn\\[3mm] 
    &\quad\times\frac{g\big(M_{\chi^0}\, \text{adj}
      \mathcal{F}\big)_{\tilde{h}^0_u \tilde{W}^3} - g'\big(M_{\chi^0}\,
      \text{adj} \mathcal{F}\big)_{\tilde{h}^0_u \tilde{B}}}{\det\mathcal{F}} + \text{
      transpose}\,. 
\end{align}
We may now employ the identity,
\begin{equation}
   \label{eq:id-mM}
f(\mM^\dagger\mM)\,\mM^\dagger\ =\ \mM^\dagger\, f(\mM\mM^\dagger)\,,
\end{equation}
which holds for an arbitrary regular function $f(x)$, in order to
rewrite~\eqref{SoftInt1} as follows:
\begin{align}
   \label{SusyInt1'}
i\Sigma^{\text{[3(a)]}}_M =\
  &\frac{1}{v_u}\mathbf{M}^*b^*_\nu\mM^\dagger\int\!\frac{d^d
    k}{(2\pi)^d}\frac{g\big(M_{\chi^0}\, \text{adj}
    \mathcal{F}\big)_{\tilde{h}^0_u \tilde{W}^3} - g'\big(M_{\chi^0}\,
    \text{adj} \mathcal{F}\big)_{\tilde{h}^0_u
    \tilde{B}}}{\det\mathcal{F}\,(k^2 - \tilde{H}^*_1)(k^2\Id_{n_R} -
    \Hdt)^2}\mD^{\sf T} 
    + \text{ transpose}\nonumber\\[3mm] 
    =\ &\mathbf{M}^*b_\nu^*\mM^\dagger\int\!\frac{d^d
       k}{(2\pi)^d}\frac{Ak^6 + Bk^4 + Ck^2 + D}{\det\mathcal{F}\,(k^2\ -
       \tilde{H}^*_1)(k^2\Id_{n_R} - \Hdt)^2}\mD^{\sf T} 
    + \text{ transpose}\,.
\end{align}
Following the same procedure as before, we eventually arrive at
\begin{align}
    \label{SoftContribution1}
i\Sigma^{\text{[3(a)]}}_M =\ &b^*_\nu\,\mD\,\Big(\Anu -
                                     \mu^*\cot\beta\Id_{n_R}\Big)\,
                                     \mM^\dagger\,\bigg[\,A\,
                                     I_4\left(\tilde{H}_1^*,\Hdt,\Hdt,
                                     \mIV^2\right)\nonumber\\[3mm] 
    &+ \mathcal{B}_{A,B}\, I_5\left(\tilde{H}_1^*,\Hdt,\Hdt,m^2_{\chi^0_{3}},\mIV^2\right)
    +
      \mathcal{C}_{A,B,C}\,I_6\left(\tilde{H}_1^*,\Hdt,\Hdt,m^2_{\chi^0_{2}},\mIII^2,\mIV^2\right)
\nonumber\\[3mm] 
    &+ \mathcal{D}_{A,B,C,D}\,I_7\left(\tilde{H}_1^*,\Hdt,\Hdt,
      \mI^2,\mII^2,\mIII^2,\mIV^2\right)\bigg]\,\mD^{\sf T}\nn\\[3mm]
    &+ \text{ transpose}\, .
\end{align}
Here, the index-valued functions $\mathcal{B}_{A,B}$, $\mathcal{C}_{A,B,C}$
and $\mathcal{D}_{A,B,C,D}$ are given in~\eqref{eq:calBCD}.

The diagram in Fig.~\ref{SoftBreakingDiagrams1}(b) can be evaluated in
a similar manner, involving the entries $\tilde{B}\tilde{B}$ and
$\tilde{W}^3\tilde{W}^3$ of the neutralino propagator matrix. More
explicitly, we find
\begin{align}
   \label{SoftContribution2} 
    i\Sigma^{\text{[3(b)]}}_M =\ &\mD\, \Big(\Anu -
  \mu^*\cot\beta\Id_{n_R}\Big)\,\mM^\dagger\,
\bigg[A' I_4\left(\tilde{H}_1,\tilde{H}^*_1, \Hdt, \mIV^2\right)\nonumber\\[2mm]
    &+ \mathcal{B}_{A',B'}\, I_5\left(\tilde{H}_1,\tilde{H}^*_1, \Hdt,
      \mIII^2,\mIV^2 \right) +
      \mathcal{C}_{A',B',C'}\,I_6\left(\tilde{H}_1,\tilde{H}^*_1,
      \Hdt,\mI^2,\mIII^2,\mIV^2\right)\nonumber\\[2mm] 
    &+ \mathcal{D}_{A',B',C',D'}\,I_7\left(\tilde{H}_1,\tilde{H}^*_1,
      \Hdt, \mI^2,\mII^2,\mIII^2,\mIV^2\right)\bigg]\,\mD^{\sf T}\nonumber\\[2mm]
    &+\,\text{transpose}\,,
\end{align}
where the index-valued functions $\mathcal{B}_{A',B'}$, $\mathcal{C}_{A',B',C'}$
and $\mathcal{D}_{A',B',C',D'}$ are determined through~\eqref{eq:calBCD},
and the coefficients $A'$, $B'$, $C'$, and $D'$ are computed 
in~\eqref{coefs'}. 

\vfill\eject

The remaining Feynman-diagrammatic contributions shown in
Figs.~\ref{SoftBreakingDiagrams1}(c)--(e) are respectively given by
\begin{align}
    \label{SoftContribution3} 
i\Sigma^{\text{[3(c)]}}_M =\ &b^*_\nu \mD\,\Big(\Anu -
                               \mu^*\cot\beta\Id_{n_R}\Big)\,
\mM^\dagger\,\bigg[A' I_5\left(\tilde{H}_1,\tilde{H}^*_1, \Hdt, \Hdt,
                               \mIV^2\right)\nonumber\\[2mm] 
    &+\mathcal{B}_{A',B'}\, I_6\left(\tilde{H}_1,\tilde{H}^*_1, \Hdt,
      \Hdt, \mIII^2,\mIV^2 \right) +
      \mathcal{C}_{A',B',C'}\,I_7\left(\tilde{H}_1,\tilde{H}^*_1,
      \Hdt, \Hdt^,\mI^2,\mIII^2,\mIV^2\right)\nonumber\\[2mm] 
    &+ \mathcal{D}_{A',B',C',D'}\,I_8\left(\tilde{H}_1,\tilde{H}^*_1,
      \Hdt, \Hdt,
      \mI^2,\mII^2,\mIII^2,\mIV^2\right)\bigg]\,\Big(\Anu^{\sf T} -
      \mu^*\cot\beta\Id_{n_R}\Big)\,\mD^{\sf
      T},\\[4mm]
    \label{SoftContribution4}
i\Sigma^{\text{[3(d)]}}_M =\ 
&b_\nu\,\mD\mM^\dagger\mM\bigg[\,A' I_5\left(\tilde{H}_1,\tilde{H}^*_1,
  \Hdt^{\sf T}, \Hdt^{\sf T}, \mIV^2\right)\nonumber\\[2mm] 
    &+\mathcal{B}_{A',B'}\, I_6\left(\tilde{H}_1,\tilde{H}^*_1, \Hdt^{\sf T}, \Hdt^{\sf T}, \mIII^2,\mIV^2 \right) +
      \mathcal{C}_{A',B',C'}\,I_7\left(\tilde{H}_1,\tilde{H}^*_1,
      \Hdt^{\sf T}, \Hdt^{\sf
      T},\mI^2,\mIII^2,\mIV^2\right)\nonumber\\[2mm] 
    &+ \mathcal{D}_{A',B',C',D'}\,I_8\left(\tilde{H}_1,\tilde{H}^*_1,
      \Hdt^{\sf T}, \Hdt^{\sf T},
      \mI^2,\mII^2,\mIII^2,\mIV^2\right)\bigg]\,\mM^\dagger\mD^{\sf T}\,,\\[4mm]
   \label{SoftContribution5}
i\Sigma^{\text{[3(e)]}}_M =\ & b^*_\nu\,\mD\mM^\dagger\bigg[\, A''
                                  I_3\left(\Hdt,\Hdt,
                                  \mIV^2\right) +
                                  \mathcal{B}_{A'',B''}\,
                                  I_4\left(\Hdt, \Hdt,
                                  \mIII^2,\mIV^2
                                  \right)\nonumber\\[2mm] 
    &+ \mathcal{C}_{A'',B'',C''}\,I_5\left(\Hdt,\Hdt,\mI^2,\mIII^2,\mIV^2\right)\nn\\[2ex] &+
                                                 \mathcal{D}_{A'',B'',C'',D''}\,I_6\left(\Hdt, \Hdt,
                                                 \mI^2,\mII^2,\mIII^2,\mIV^2\right)\bigg]\mD^{\sf
                                                 T}\, . 
\end{align}
where the coefficients that $A''$, $B''$, $C''$ and $D''$ are given in
(\ref{coefs''}). 

We conclude this section by making a few important remarks.  First, we
note that the flavour structure of the self-energy contribution given
in~\eqref{SoftContribution2} depends strongly on the texture of
$\Anu$, and it is independent of the sneutrino Majorana mass matrix
$\bnu$. Second, we observe that the last self-energy contribution
in~\eqref{SoftContribution5} becomes rather relevant since its flavour
structure is quite similar to that obtained by the $\nu_R$SM graphs
in Fig.~\ref{SMdiagrams}. It can be roughly of the same order without
assuming very high values for the parameter $b_\nu$. This allows the
existence of a parameter space where the $\nu_R$SM contributions are
screened by this diagram. This fact will be used in the next section
for defining appropriate benchmark scenarios for radiative neutrino
masses in the $\nu_R$MSSM.  Finally, the diagram in
Fig.~\ref{SoftBreakingDiagrams1}(e) is non-zero, even if all soft
masses are set to zero, provided the $\mu$ parameter is non-zero.
To be specific, there is a SUSY counterpart to Fig.~\ref{SMdiagrams}(a) involving a single
$\mu$-dependent chirality flip on the higgsino line in the
loop. However, this graph contributes 
with four powers in the neutrino Yukawa matrix~$\Y_\nu$, and so
it can be consistently ignored in the leading second-order approximation
of~$\Y_\nu$ that we have been working here.

\vfill\eject

\section{Numerical Results}\label{sec:Results}

We will now perform a numerical analysis for a few representative
scenarios that give rise to radiative neutrino masses in the
$\nu_R$MSSM. As there are many parameters that can vary independently,
we choose the baseline benchmark model exhibited in
Table~\ref{tab:Benchmark}. In addition, as discussed in the previous
section and in Appendix~\ref{AppendixHiggsSector}, we use an effective
quartic coupling $\lambda_{\text{eff}}$ for the quadrilinear
interaction $(H^\dagger_u H_u)^2$ induced by quantum effects, such
that the mass of the $h$ boson is equal to that of the observed
SM-like Higgs boson, i.e.~$m_h = 125.38\pm0.14$~GeV.  Finally, unless stated
otherwise, we assume a universal structure for the soft SUSY-breaking
parameters, as given in~\eqref{eq:SoftU}.

\begin{table}[]
    \centering
    \begin{tabular}{|c|l|}
\hline
& \\[-3mm]
Parameter & Numerical value/interval\\[2mm]
\hline\hline
& \\[-3mm]
        $\tan\beta$ & ~2, 20\\
       $\mu$ & ~1200 GeV\\
        $M_1$ & ~1500 GeV\\
        $M_2$ & ~1500 GeV\\
        $m_h$ & ~125.38$\pm$0.14 GeV\\
        $m_A$ & ~5000 GeV\\
        $m_H$ & ~5002.8 GeV\\
        $m_N$ & ~500 GeV\\
        $\msqL$ & ~(3500 GeV)$^2$\\
        $\mu_R$ & ~$[10^{-6},\,10^2]$ GeV\\[2mm]
\hline
      \end{tabular}
    \caption{Input parameters for our baseline benchmark scenario in
      the $\nu_R$MSSM.}
    \label{tab:Benchmark}
\end{table}

An important constraint in our numerical analysis is the compatibility
of the effective neutrino mass matrix $\M_{\nu_L} = \Sigma_M(0)$, obtained through
\eqref{NuMassMatrix}, with the low-energy neutrino data. To be
explicit, we require that 
\begin{align}
   \label{PMNS}
\Sigma_M(0)\: =\: \mathbf{M}^{\rm exp}_{\nu_L}\: =\: U^{\sf
  T}\,\wh{\mathbf{M}}_{\nu_L}U\,,
\end{align}
where $U$ is the PMNS lepton mixing matrix~\cite{Pontecorvo:1957qd,Maki:1962mu}, and
$\wh{\mathbf{M}}_{\nu_L} = \text{diag}(m_1, m_2, m_3)$ and $m_{1,2,3}$
are the light neutrino masses. To match our self-energy contributions
$\Sigma_M$ derived in the previous section to
$\mathbf{M}^{\rm exp}_{\nu_L}$, we neglect non-unitarity effects that
come from light-to-heavy neutrino mixing and assume that the charged
lepton Yukawa matrix ${\bf Y}_l$ is diagonal and already expressed in the mass
basis. With the above assumptions in mind, the matrix $U$ can be
parametrised as follows:
\begin{equation}
U\ =\ \begin{pmatrix}
c_{12}c_{13} & s_{12}c_{13} & s_{13}e^{-i\delta} \\
-s_{12}c_{23}-c_{12}s_{23}s_{13}e^{i\delta} & c_{12}c_{23}-s_{12}s_{23}s_{13}e^{i\delta} & s_{23}c_{13}\\
s_{12}c_{23}-c_{12}c_{23}s_{13}e^{i\delta} & -c_{12}s_{23}-s_{12}c_{23}s_{13}e^{i\delta} & c_{23}c_{13}
\end{pmatrix}\times \text{diag}(e^{i\alpha_1/2},e^{i\alpha_2/2},1),
\end{equation}
where $c_{ij} = \cos\theta_{ij}$ and $s_{ij} = \sin\theta_{ij}$ are
the neutrino mixing angles, $\delta$ is the so-called Dirac  phase, and
$\alpha_{1,2}$ are the Majorana phases. 

In our numerical estimates, we assume normal neutrino mass ordering.
This choice is favoured by recent global fits of neutrino oscillation
data. For definiteness, we use the latest best fit values for the neutrino
oscillation parameters~\cite{deSalas:2020pgw}:
\begin{align}
    &\Delta m^2_{21}\equiv m^2_2 - m^2_1 = 7.50\times
      10^{-5}\,(\text{eV})^2,\quad\Delta m^2_{31}\equiv m^2_3 - m^2_1
      = 2.56\times 10^{-3} \,(\text{eV})^2,\\[2mm] 
    &\sin^2\theta_{12} = 34.3^{\circ},\quad \sin^2\theta_{23} =
      48.79^{\circ} ,\quad\sin^2\theta_{12} = 8.58^{\circ},\quad
      \delta = 216^{\circ}. 
\end{align}
Furthermore, we set $m_1 = 0$ and $\alpha_{1,2} = 0$. 

Let us first consider the radiative ISS scenario of the $\nu_R$MSSM
described in Section~\ref{sec:nuRMSSM},
with~${\boldsymbol{\mu}_R \neq {\bf 0}_n}$, $\boldsymbol{\mu}_S = {\bf 0}_n$, and
$n=2$ pairs of singlet neutrinos.  This model necessarily implies a
massless neutrino, i.e.~$m_1 = 0$, at the one-loop level. This should
be contrasted to the standard ISS model, with $\boldsymbol{\mu}_R = 0$ and
$\boldsymbol{\mu}_S \neq 0$, in which case one pair of singlet
neutrinos (i.e.~$n=1$) would have been sufficient to accommodate the neutrino
oscillation data (see, e.g.~\cite{Hirsch:2009ra}).

For our illustrative purposes, we will initially assume that the
right-handed sneutrino sector is supersymmetric and only consider
minimal departures from it, by taking that either $b_\nu$ {\em or} $A_\nu$ 
is non-zero, but putting $m^2_{\tilde{\nu}} = 0$ in all settings. For such
scenarios, the dominant contributions to the effective neutrino mass
matrix $\M_{\nu_L}$  arise from the diagrams shown in
Figs.~\ref{SMdiagrams}, \ref{Supersymmetric graphs} and
\ref{SoftBreakingDiagrams1}(b). For the latter graph, the dominant
SUSY effect on $\M_{\nu_L}$ comes from the $\mu$ term. 
In fact, this last effect can screen completely that $\nu_R$SM contribution
from the graphs in Fig.~\ref{SMdiagrams}, even in the limit of
a fully supersymmetric neutrino sector with $b_\nu = A_\nu = 0$. 

\begin{figure}[t]
\includegraphics[scale=0.53]{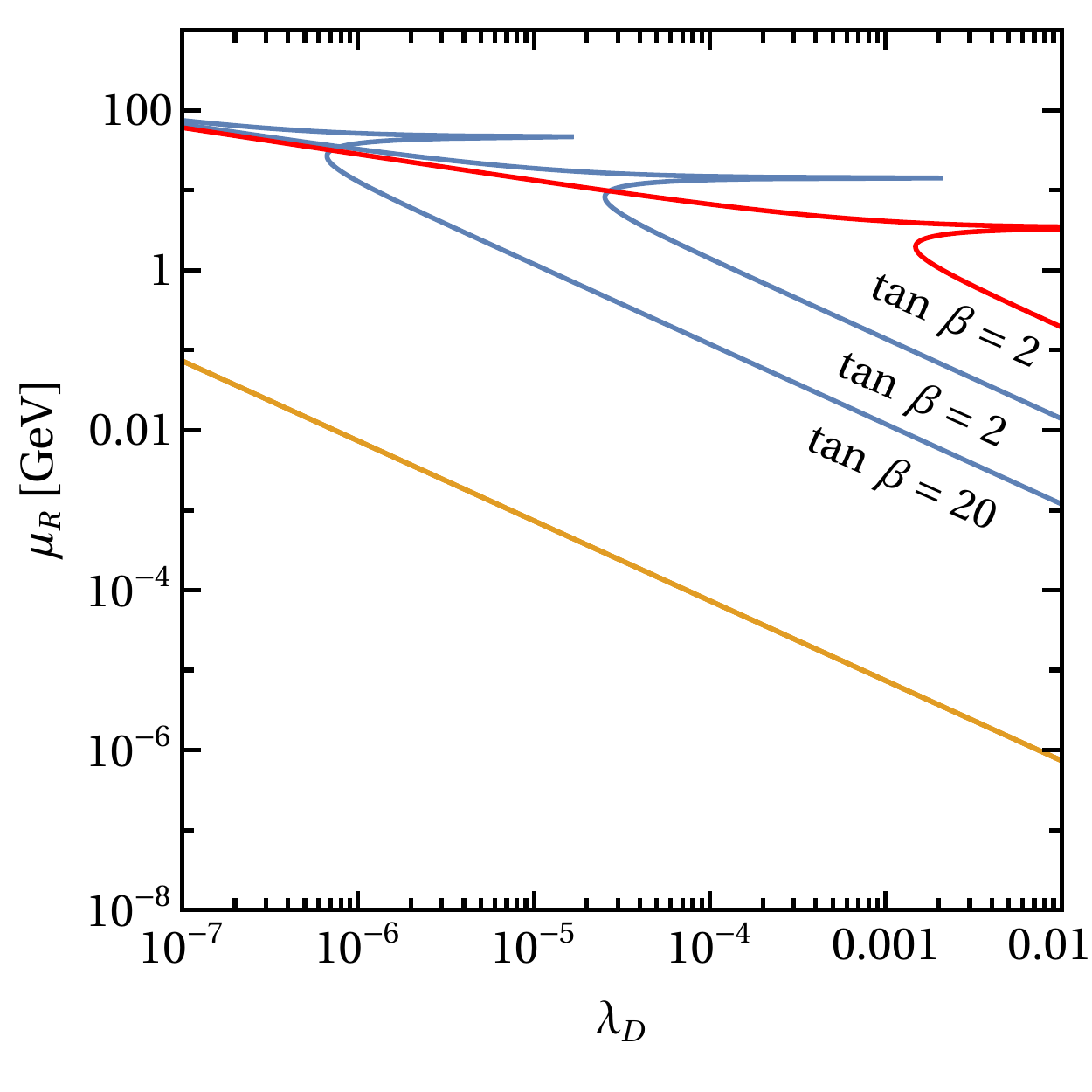}
\hspace{5mm}
\includegraphics[scale=0.53]{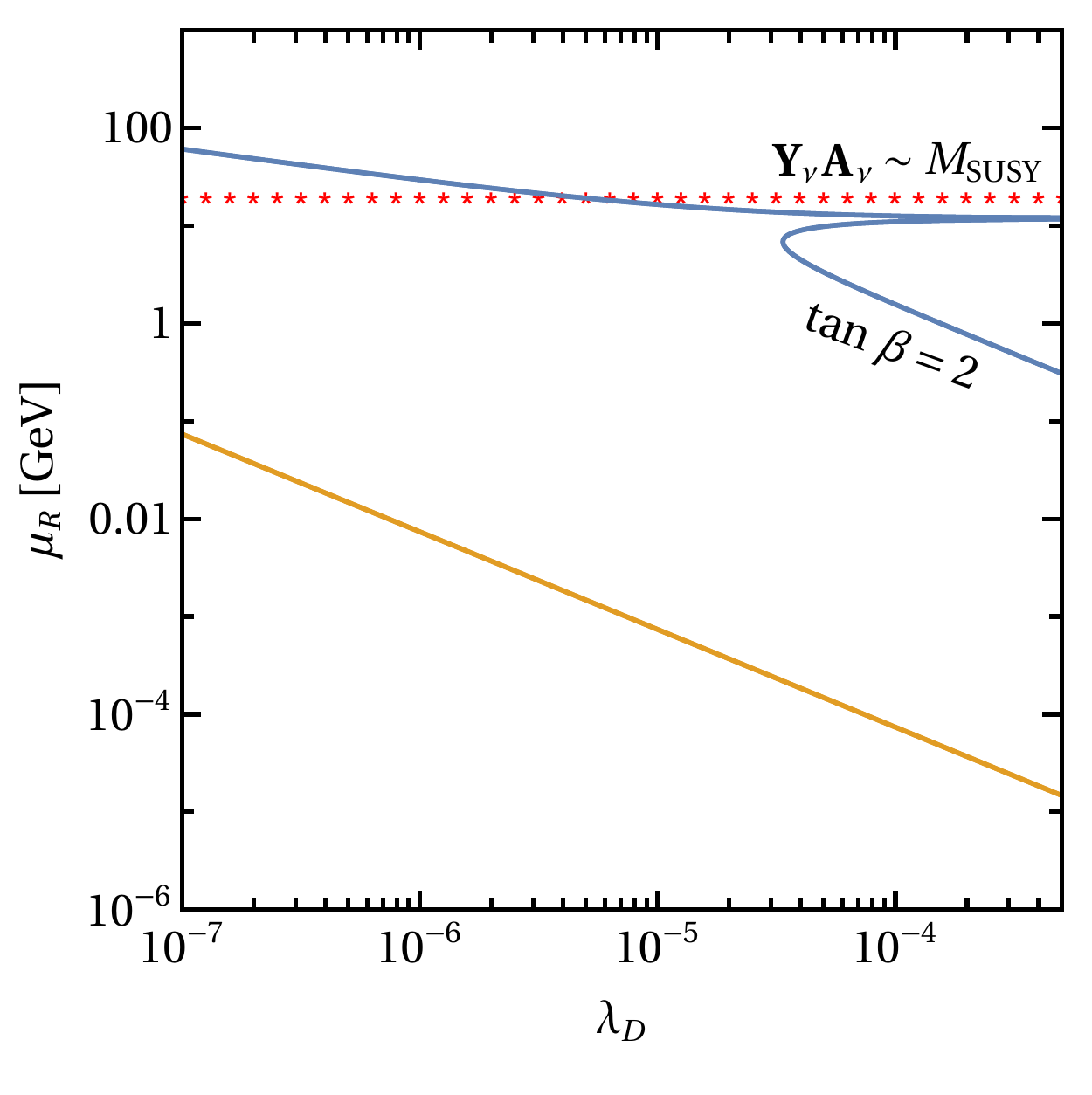}
\includegraphics[scale=0.75]{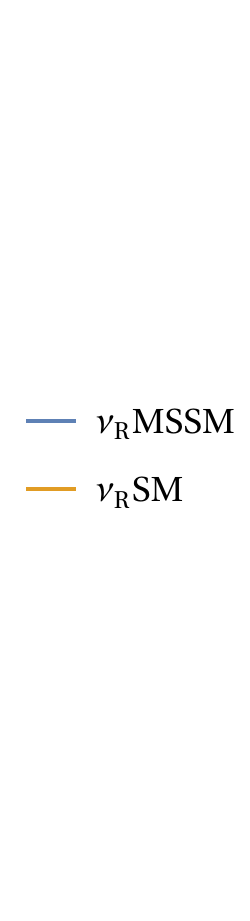}
	\caption{Numerical estimates of the LNV parameter $\mu_R$
          versus $\lambda_D$ for two scenarios: (i)~$b_\nu\neq 0$,
          with $A_\nu = 0$ (left frame) and (ii)~$A_\nu\neq 0$, with
          $b_\nu = 0$ (right frame).  For scenario~(i), the screening
          values for the soft bilinear parameter are: $b^{0}_\nu =
          0.289$~GeV for $\tan\beta=2$ and $b^{0}_\nu =
          0.399$~GeV for  $\tan\beta=20$. For~scenario~(ii),
          the screening soft trilinear coupling is:
          $A^{\text{0}}_\nu \simeq 389.1$ TeV. The line in red in the
          left panel shows the predictions obtained from a fully
          supersymmetric singlet sector, achievable for
          $\mu = \mu^0\simeq-770.3$ TeV. The horizontal dashed line
          on the right frame indicates the value of $\mu_R$, at
          which the largest entry of $\Y_\nu \Anu$ reaches
          $M_{\text{SUSY}}$.}
	\label{muR_vs_lamD}
\end{figure}

Following~\cite{Dev:2012sg}, we define the parameter
$m^{\rm max}_D = \text{max}|{({\bf M}_D)}_{ij}|$, which enables us to
introduce the quantity
\begin{equation}
   \label{eq:lambdaD}
\lambda_D\ =\ (m^{\rm max}_D)^2/m^2_N\,.
\end{equation}
Notice that $\lambda_D$ quantifies the size of the light-to-heavy
neutrino mixing in quadrature.

In Fig.~\ref{muR_vs_lamD}, we present exclusion plots of the SUSY
parameter $\mu_R$ as a function of~$\lambda_D$, for two scenarios:
(i)~left panel:~$b_\nu\neq 0$ ($A_\nu = 0$) and (ii)~right panel:
$A_\nu\neq 0$ ($b_\nu = 0$). In the left panel of
Fig.~\ref{muR_vs_lamD}, we see an effect of complete cancellation for
$b_\nu = b^{0}_\nu = 0.289$~GeV, when $\tan\beta=2$, and
$b_\nu = b^{0}_\nu = 0.399$~GeV, when $\tan\beta=20$. Also, we see a
similar effect to occur in the right panel of Fig.~\ref{muR_vs_lamD}, for
$A_\nu = A^0_\nu \simeq 389.1$~TeV. It is interesting to notice that
for large regions of the parameters $b_\nu$ and $A_\nu$, the allowed
values for the LNV parameter $\mu_R$ can be nearly four orders of
magnitude higher in the $\nu_R$MSSM than in the $\nu_R$SM. In fact,
the soft SUSY-screening phenomenon can be so strong, that the LNV
scale $\mu_R$ can get close to the electroweak scale for relatively
large light-to-heavy neutrino mixings of order $10^{-2}$,
corresponding to $\lambda_D \sim 10^{-4}$. Most interestingly, as can
be seen from the left panel of Fig.~\ref{muR_vs_lamD} given by the
line in red, such a SUSY-screening phenomenon can take place for an
exact supersymmetric sneutrino sector, with $b_\nu = A_\nu = 0$, but
for an unusually large value of the $\mu$ parameter,
$\mu = \mu^0 \simeq -770.3$~TeV. This very large value of $\mu$ would
require superheavy squark masses of order $10^3$~TeV and higher, as
could happen in scenarios of
split~SUSY~\cite{Wells:2003tf,ArkaniHamed:2004fb, Giudice:2004tc}.

\begin{figure}[t]
\includegraphics[scale=0.53]{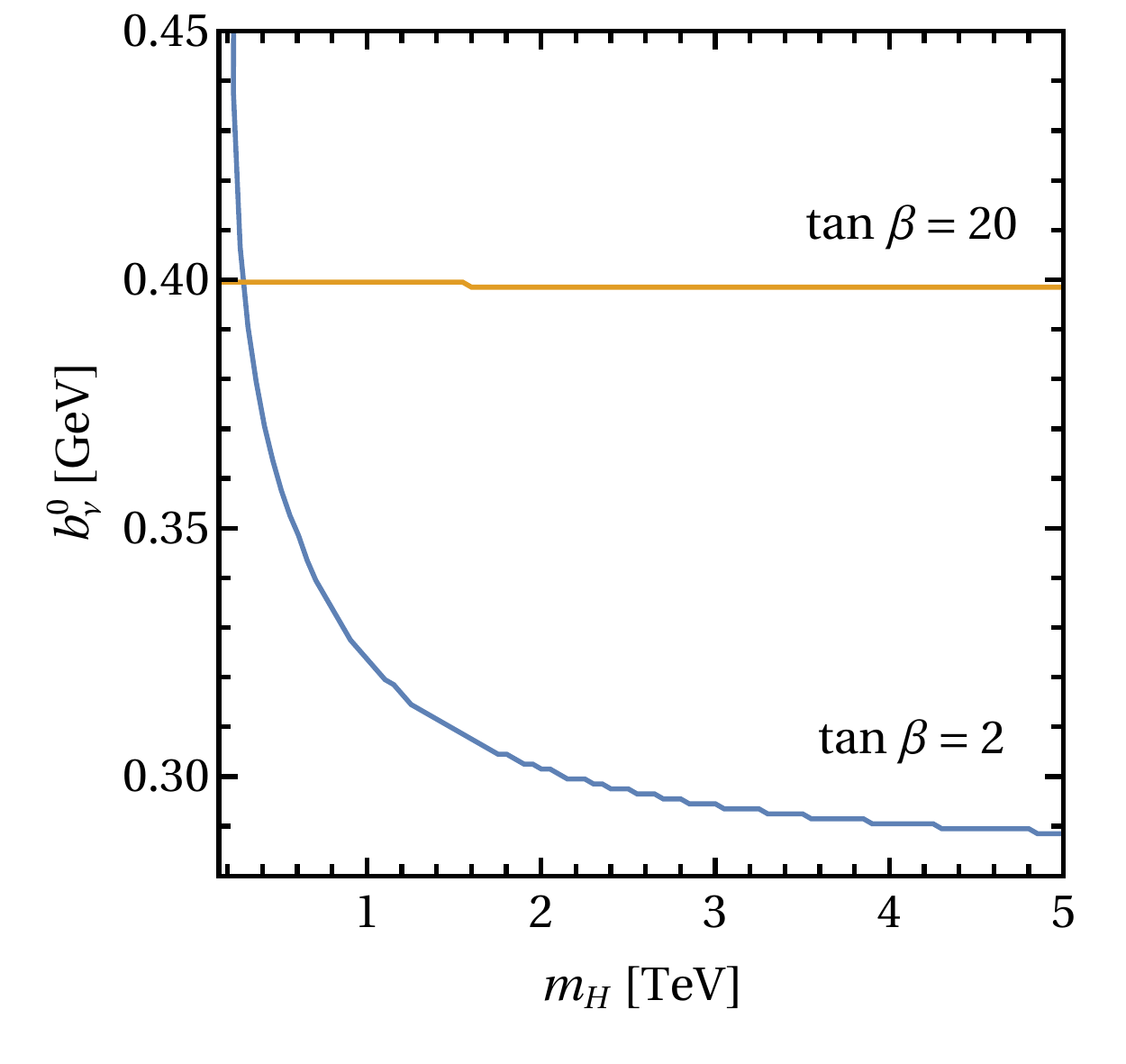}
\hspace{5mm}
\includegraphics[scale=0.53]{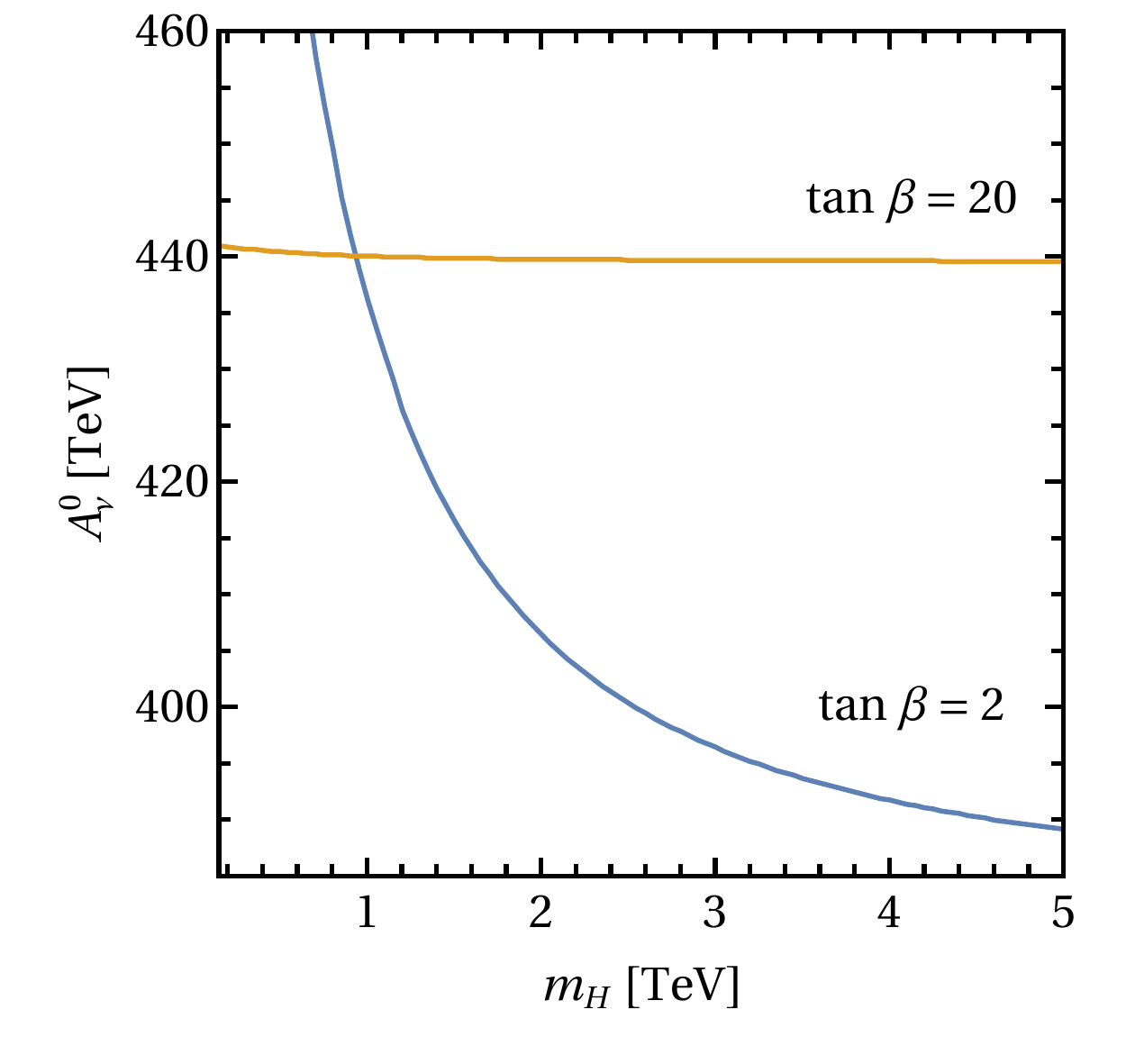}
	\caption{The dependence of the exact screening values for
          $b^0_\nu$ (left frame) and $A^0_\nu$ (right frame) on the
          heavy Higgs mass $m_H$, after setting $\mu_R = 10^{-2}$~GeV.}
	\label{cancellation_values_vs_mH}
\end{figure}

In Fig.~\ref{cancellation_values_vs_mH}, we display the dependence of
the cancelling values $b^0_\nu$ and $A^0_\nu$ as functions of the
heavy CP-even Higgs-boson mass $m_H$.  As can be seen from the
left panel in Fig.~\ref{cancellation_values_vs_mH}, the $\nu_R$SM-like
contribution decreases somewhat as the heavy scalar sector decouples,
and as such,  a smaller value of $b^0_\nu$ will be needed to provide the required
cancellation. At the larger value of~$\tan\beta = 20$, the quantum loop
effects on the quartic coupling $(H^\dagger_u H_u)^2$ become less
pronounced, thus making the diagram in Fig.~\ref{SMdiagrams}(a)
less significant. The same features can be seen in the right panel
of  Fig.~\ref{cancellation_values_vs_mH}, where the dependence of
$A^0_\nu$ on $m_H$ is exhibited. 

The above SUSY-screening phenomenon may prevent specific flavour
structures of $\Y_\nu$ that may occur in the $\nu_R$SM from producing
too large finite quantum corrections to light neutrino masses, when
the heavy singlet neutrinos happen to be non-degenerate in
mass~\cite{Pilaftsis:1991ug}. As an illustrative example, let us
consider a three-generation scenario with $n_R =3$ right-handed
neutrinos.  After SSB, the Dirac mass matrix of such a scenario takes
on the form
\begin{equation}
  \label{eq:mD}
{\bf m}_D \ =\ \frac{v_u}{\sqrt{2}}\left( \begin{array}{ccc}
a &  b\, e^{2i\pi/3} & c\, e^{-2i\pi/3}\\
a &  b\, e^{2i\pi/3} & c\, e^{-2i\pi/3}\\
a &  b\, e^{2i\pi/3} & c\, e^{-2i\pi/3}
\end{array} \right)\; ,
\end{equation}
when expressed in a flavour basis in which the singlet neutrino mass matrix is diagonal and
positive, i.e.
\begin{equation}
{\bf m}_M\ =\ \text{diag}\,\Big( m_N\,,\, m_N + \delta_N\,,\, m_N + 2\delta_N\Big)\,.
\end{equation}
Here, the parameter $\delta_N$ quantifies the breaking of the mass
degeneracy in the heavy-neutrino sector. Moreover, the Yukawa
parameters $b$ and $c$ are not independent of $a$, but they obey
the relations:
\begin{equation}
    \label{eq:abc}
b\ =\ a\, \sqrt{1\: +\: \frac{\delta_N}{m_N}}\;,\qquad c\ =\ a\,
\sqrt{1\: +\: \frac{2\delta_N}{m_N}}\;,
\end{equation}
such that the light neutrinos are exactly massless at the tree level,
i.e.~$\mathbf{m}_\nu = {\bf 0}_3$, as can be easily determined
from~\eqref{eq:seesaw}. Beyond the Born approximation, such a scenario
leads to too large radiative neutrino masses in the
$\nu_R$SM~\cite{Pilaftsis:1991ug}, in conflict with neutrino
oscillation data, unless $\delta_N/m_N \stackrel{<}{{}_\sim} 10^{-5}$,
for $m_N = 500$~GeV and light-to-heavy neutrino
mixing~$({\bf m}_D)_{11}/m_N = 10^{-2}$. However, in the $\nu_R$MSSM,
we find that this constraint may be relaxed drastically by several
orders of magnitude.

\begin{figure}[t]
	\includegraphics[scale=0.55]{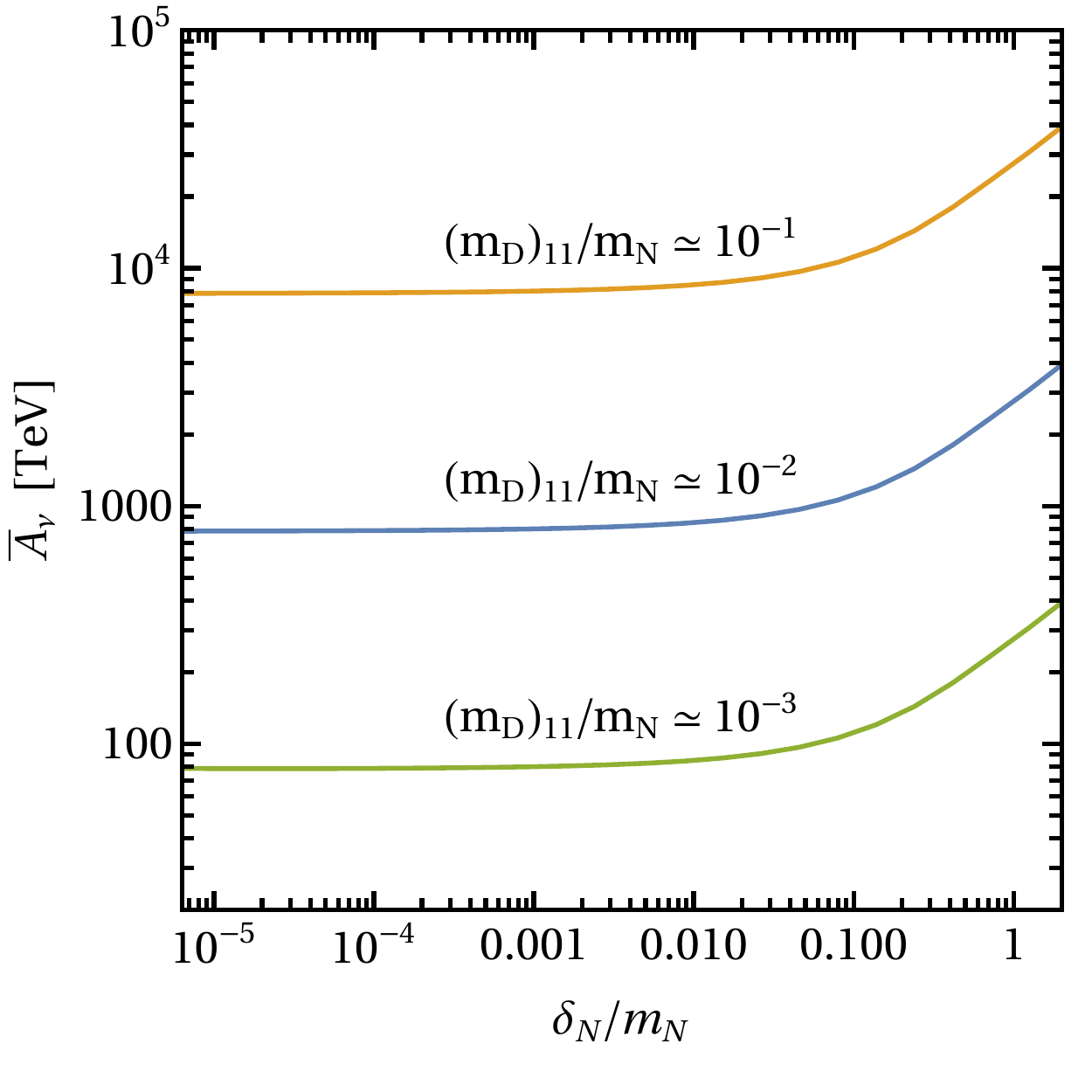}
	\caption{The dependence of $\overline{A}_\nu$
          [cf.~\eqref{eq:barAnu}] on the LNV parameter $\delta_N/m_N$,
          for $a = 0.5,\ 0.05,\ 0.005$ as shown from the upper to
          lower lines, respectively. The MSSM parameters are the same
          as in Table~\ref{tab:Benchmark} (with $\tan \beta = 2$),
          while the mass of the lightest neutrino is set to zero.}
	\label{figure:Anu11_vs_deltaN/mN}
\end{figure}

To showcase this relaxation of the constraint on the LNV parameter
$\delta_N$, we consider a flavour scenario in which $\Anu$ has its own
flavour structure that is independent of the neutrino Yukawa matrix
${\bf Y}_\nu$ as this can be inferred from~\eqref{eq:mD}.  In this
case, it is possible to find an appropriate form for~$\Anu$ and a
value for~$b_\nu$, so as to reproduce the observed neutrino mass
matrix $\mathbf{M}^{\rm exp}_{\nu_L}$ given in~\eqref{PMNS}. This is
illustrated in Fig. \ref{figure:Anu11_vs_deltaN/mN}, which shows the
dependence of the {\em average} norm of $\Anu$, defined as
\begin{equation}
    \label{eq:barAnu}
\overline{A}_\nu\, \equiv\,
\frac{1}{3}\,\text{Tr}^{1/2}\left(\Anu\Anu^\dagger\right)\, ,
\end{equation}
on the LNV dimensionless parameter $\delta_N/m_N$, for different
values of light-to-heavy neutrino mixing. Interestingly enough, it can
be seen that large heavy neutrino mass differences 
$\delta_N/m_N \sim 10^{-1}$, with sizeable light-to-heavy neutrino
mixing $({\bf m}_D)_{11}/m_N \sim 10^{-2}$, are allowed, provided that the
trilinear parameter $\overline{A}_\nu$ is sufficiently large, i.e.
$\overline{A}_\nu \approx 1000$~TeV.

In summary, our numerical estimates have revealed that larger values
of LNV mass parameters${}$, such as~$\mu_R$ and $\delta_N$, are
allowed in the $\nu_R$MSSM than in the $\nu_R$SM, while the squared
light-to-heavy neutrino mixing as measured by $\lambda_D$ or
$({\bf m}_D)_{11}/m_N$ can be equally sizeable.  Remarkably enough, this
finding implies\- that signatures of LNV mediated by non-degenerate
heavy Majorana neutrinos could be on the verge of being discovered
with the upcoming LHC data~\cite{Deppisch:2015qwa}.

\section{Discussion}\label{sec:Conclusions}

We have studied supersymmetric scenarios of radiative neutrino masses
that may occur in the Minimal Supersymmetric Standard Model with
low-scale right-handed neutrinos ($\nu_R$MSSM).  To~have good
control of the impact of the SUSY non-renormalization theorems on the
loop-induced neutrino masses, we have carefully performed all the
Feynman-diagrammatic computations in the weak and flavour bases,
rather in the mass basis. In this way, we have been able to identify a
new mechanism for naturally suppressing the light neutrino masses
beyond the traditional seesaw paradigm. In the context of the {\em
  radiative} inverse seesaw scenario first introduced and studied
in~\cite{Dev:2012sg}, the smallness of the observed light neutrino
masses may be the result of a soft SUSY-screening effect from a nearly
supersymmetric singlet neutrino sector. An important consequence of
this effect is that unlike in the {\em non}-supersymmetric scenario
of~\cite{Dev:2012sg}, the singlet seesaw scale~$m_N$ and the LNV scale
$\mu_R$ can now be both  comparable in size, e.g.~of the electroweak
order, whilst the size of the light-to-heavy neutrino mixing squared,
$\lambda_D$, can be enhanced up to the~$10^{-2}$
level~(cf.~Fig.~\ref{muR_vs_lamD}).

We note that the radiative generation of the light neutrino masses
involves almost the full spectrum of particles of the $\nu_R$MSSM.
Specifically, the inclusion of third generation quarks and scalar
quarks at the two-loop level are important to obtain an effective
quartic Higgs coupling compatible with the SM Higgs-boson mass.  If
$R$-parity is conserved, the lightest right-handed sneutrino can be a
successful Dark Matter candidate~\cite{Gopalakrishna:2006kr}.  Hence,
both the visible and DM sectors contribute, through one and higher
loops, in order to generate the observed light neutrino
masses. As~such, the radiative screening mechanism 
presented in this paper is a consequence of {\em pangenesis}, as it
requires almost the entire field content of the theory in order to be realised.

In this pangenic framework of the $\nu_R$MSSM, the strict
constraints~\cite{Kersten:2007vk,Deppisch:2015qwa} from light neutrino
masses on LNV signatures from heavy Majorana neutrinos can be relaxed
significantly, or even eliminated.  On the other hand, possible
observation of LNV signatures mediated by pairs of heavy Majorana
neutrinos with masses $\sim (m_N\pm \mu_R)$ at high-energy colliders
would give rise to renewed impetus in searches for supersymmetric
right-handed sneutrinos in a very similar and highly correlated mass
range.

Besides the above new aspects of collider phenomenology, minimal
extensions of the $\nu_R$MSSM not only can solve the infamous
$\mu$-problem, but they can have profound cosmological implications as
well, explaining the observed nearly scale-invariant cosmic microwave
background spectrum and the flatness problem of the Universe through
inflationary dynamics~\cite{Garbrecht:2006az}. In such extensions,
right-handed sneutrinos can become {\em thermal} DM
particles~\cite{Deppisch:2008bp}.  Therefore, in light of the recent
laboratory and cosmological data, it would be interesting to perform
an updated, fully fledged analysis of such minimal extensions of the
$\nu_R$MSSM.

\bigskip

\section*{Acknowledgements} 
\vspace{-3mm}

\noindent
The work of AP is supported in part by the
Lancaster--Manchester--Sheffield Consortium for Fundamental\- Physics,
under STFC Research Grant No. ST/P000800/1. The work of PCdS is funded
by Becas Chile, ANID-PCHA/2018/72190359. The Feynman diagrams shown in
this article\- were generated with the TikZ-Feynman package
\cite{Ellis:2016jkw}.

\vfill\eject

\appendix

\section{The Higgs Sector of the {\boldmath $\nu_R$}MSSM at Tree 
  Level}\label{AppendixHiggsSector} 

Here we briefly review the SUSY limit in the $\nu_R$MSSM, while
describing our conventions for its Higgs sector. In fact, this sector
becomes identical to the that of the MSSM at tree level.  After SSB,
the Higgs doublets $H_u$ and $H_d$ may be linearly expanded about
their VEVs, $v_u$ and $v_d$, as follows:
\begin{align}
    H_u = \begin{pmatrix}
    H^+_u\\
    \frac{1}{\sqrt{2}}(v_u + \phi_u + ia_u)
    \end{pmatrix},\qquad
    H_d = \begin{pmatrix}
    \frac{1}{\sqrt{2}}(v_d + \phi_d + ia_d)\\
    H^-_d
    \end{pmatrix}\,.
\end{align}
In the MSSM, the electroweak symmetry breaking is in general 
connected to SUSY breaking. To see this, we start by
analyzing the neutral part of the Higgs potential. This is given by
\begin{align}
    V^0_H = &\frac{1}{8}(g'^2 + g^2)\,
\big(|H^0_d|^2 - |H^0_u|^2 \big)^2\: +\: \big(m^2_{H_d} + |\mu|^2 \big)\, |H^0_d|^2 +
              \big (m^2_{H_u} + |\mu|^2 \big)\,|H^0_u|^2\nonumber\\ 
    &+ \big(B\mu\, H^0_d H^0_u\: +\: \textrm{H.c.}\big)\, , 
\end{align}
where $H^0_{u,d}$ are the neutral components of the Higgs doublets $H_{u,d}$.
At its minimum, the scalar potential $V^0_H$ takes on the form
\begin{align}
    V^{\rm vac}_H = \frac{1}{32}(g'^2 + g^2)(v^2_d - v^2_u)^2 +
  \frac{1}{2}(m^2_{H_d} + |\mu|^2)v^2_d + \frac{1}{2}(m^2_{H_u} +
  |\mu|^2)v^2_u + \big( B\mu\, v_d v_u\: +\: \textrm{H.c.}\big)\, .
\end{align}
Assuming that $B\mu$ is real, the minimisation conditions for $V^{vac}_H$
simplify to 
\begin{align}
    m^2_{H_d} + |\mu|^2 = -B\mu\frac{v_u}{v_d} + \frac{1}{8}(g^2 + g'^2)(v^2_u - v^2_d)\,,\\
    m^2_{H_u} + |\mu|^2 = -B\mu\frac{v_u}{v_d} + \frac{1}{8}(g^2 + g'^2)(v^2_d - v^2_u)\,.
\end{align}
It is now not difficult to see that when the soft SUSY-breaking terms
are set to zero, these conditions can only be fulfilled for non-zero
VEVs if $\mu = 0$, which in turn implies that $v_u = v_d$. As a consequence, 
in the absence of any soft masses, the SUSY limit in the $\nu_R$MSSM
is attained for $\mu = 0$ and $\tan\beta \equiv v_u / v_d = 1$.

The part of the Lagrangian containing the CP-odd scalar masses is given by
\begin{align}
    \mathcal{L}^{\text{mass}}_{\text{CP-odd}} = -\frac{1}{2}\begin{pmatrix}
    a_u & a_d
    \end{pmatrix}
    \begin{pmatrix}
    B\mu \cot\beta & B\mu\\
    B\mu & B\mu \tan\beta
    \end{pmatrix}
    \begin{pmatrix}
    a_u\\
    a_d
    \end{pmatrix}\, .
\end{align}
From this last expression, we see that one mass eigenstate is massless
corresponding
to the would-be Goldstone boson $G$ to be eaten by the longitudinal 
polarization of the $Z$ boson. Instead, the second mass eigenstate $A$ 
has a non-zero squared mass given by 
\begin{align}
    m^2_A = \frac{B\mu}{\sin\beta\cos\beta}\ .  
\end{align}
On the other hand, the CP-even mass Lagrangian reads
\begin{align}
	\label{eq:CP even mass matrix}
    \mathcal{L}^{\text{mass}}_{\text{CP-even}} = -\frac{1}{2}\begin{pmatrix}
    \phi_u & \phi_d
    \end{pmatrix}
    \begin{pmatrix}
    B\mu\cot\beta + \widetilde{M}^2_Z\sin^2\beta & -B\mu - M^2_Z\cos\beta\sin\beta\\
    -B\mu - M^2_Z\cos\beta\sin\beta & B\mu\tan\beta + M^2_Z\cos^2\beta
    \end{pmatrix}
    \begin{pmatrix}
    \phi_u\\
    \phi_d
    \end{pmatrix}.
\end{align}
Note that in (\ref{eq:CP even mass matrix}), we promoted the up-type
Higgs-boson quartic coupling from its tree-level value,
$\lambda_{\rm tree} = (g^2+g'^2)/8$, to the effective coupling
$\lambda_{\rm eff}$ given in~\eqref{eq:lambda_eff}. Specifically, we
have defined the mass parameter squared:
$\widetilde{M}^2_Z\equiv 2\lambda_{\rm eff}(v^2_u + v^2_d)$, which
reduces to the standard tree-level result for~$M^2_Z$, when
$\lambda_{\rm eff}$ is replaced with $\lambda_{\rm tree}$.  Denoting
with $M^2_{\cal S}$ the $2\times2$ CP-even scalar mass matrix
described by the Lagrangian in~\eqref{eq:CP even mass matrix}, we may
compute the two mass eigenstates, often called the light and heavy
Higgs bosons, $h$ and $H$, as follows:
\begin{align}
    \label{eq:HiggsMassFormula}
    m^2_{h,H} = \frac{1}{2}\left({\rm Tr}\,M^2_{\cal S} \mp \sqrt{{(\rm Tr}\,M^2_{\cal S})^2 - 
4\det M^2_{\cal S}}\right)\,,
\end{align}
with
\begin{align}
    \label{eq:Trace CP even}
{\rm Tr}\,M^2_{\cal S}\, =\ & m^2_A + \widetilde{M}^2_Z\sin^2\beta
      + M^2_Z\cos^2\beta\,,\\
    \label{eq:Det CP even}
\det M^2_{\cal S}\, =\ & m^2_A\left(M^2_Z\cos^4\beta +
                    \widetilde{M}^2_Z\sin^4\beta\right)\: -\:
                    M^2_Z\left(2m^2_A + M^2_Z -
                    \widetilde{M}^2_Z \right)\cos^2\beta\sin^2\beta\, .
\end{align}
In our analysis, we adopt a simplified approach, where the effective
coupling $\lambda_{\rm eff}$ is chosen such that the mass $m_h$ of the
lightest CP-even scalar in the ($\nu_R$)MSSM coincides with the
corresponding one for the observed SM-like Higgs resonance at the
LHC~\cite{Sirunyan:2020xwk},
i.e.~$m_h =125.38\pm0.14$~GeV~[cf.~Table~\ref{tab:Benchmark}].

\section{Neutralino and Higgs Propagators in the Weak
  Basis}\label{AppendixPropagators} 

Here, we derive the analytic matrix structure of the neutralino
propagator in the weak basis. To start with, we first quote the
neutralino mass matrix in the basis
$(\tilde{B},\tilde{W}^3,\tilde{h}^0_u,\tilde{h}^0_d)^{\sf T}$, i.e.
\begin{equation}
M_{\chi^0} = \begin{pmatrix}
    M_1 & 0 & \frac{1}{2} g' v_u & -\frac{1}{2} g' v_d \\
    0 & M_2 & -\frac{1}{2} g v_u & \frac{1}{2} g v_d \\
    \frac{1}{2} g' v_u & -\frac{1}{2}g v_u & 0 & -\mu  \\
    -\frac{1}{2} g' v_d & \frac{1}{2} g v_d & -\mu  & 0 
\end{pmatrix}\, .
\end{equation}
Then, in terms of~$M_{\chi^0}$, we define the $4\times 4$-dimensional matrix
\begin{equation}
   \label{eq:calF}
    \mathcal{F}(k)\ =\ k^2 \Id_4 - M_{\chi^0}^* M_{\chi^0}\ .
\end{equation}
We can prove that the left chiral component of the tree-level
neutralino propagator can be written as
\begin{align}
    P_L G^{(2)}_{\chi^0}(k) P_L = i M_{\chi^0} \mathcal{F}(k)^{-1}P_L.
\end{align}
Employing the linear algebra relation,
\begin{align}
    A^{-1} = \frac{1}{\det A}\:\text{adj}A\, ,
\end{align}
which is valid for any invertible matrix $A$ with its adjunct denoted 
as~$\text{adj}A$,  we get
\begin{align}
    P_L G^{(2)}_{\chi^0}(k) P_L &= \frac{i}{\det\mathcal{F}}
                                  M_{\chi^0}\:
                                  \text{adj}\mathcal{F}\, P_L\nonumber\\ 
    &= \frac{i}{\prod^4_{i=1}\left(k^2 -
      m^2_{\chi^0_i}\right)}M_{\chi^0}\:
      \text{adj}\mathcal{F}\, P_L\, .\label{NeutralinoPropa} 
\end{align}
This expression proves very convenient when performing the Feynman
parametrisation of the loop integrals that involve this propagator. 

Likewise, the U(1)$_Y$-violating part of the $H_u$ propagator can be obtained 
after inverting the expression,  
\begin{align}
    \Gamma^{(2)}_H(p)\ =\ p^2 \Id_4 - M^2_H\, ,
\end{align}
and evaluating the respective entry,  
where  $M^2_H$ is the Higgs mass matrix expressed in the weak basis,
$(H^0_d,H^0_u,H^{0*}_d,H^{0*}_u)^{\sf T}$. In this way, we may derive 
\begin{align}
	\label{eq:Hu propagator}
    G^{(2)}_{H^{0*}_u H^{0*}_u}(p) = \frac{i\widetilde{M}^2_Z\sin^2\beta(p^2 +
  m^2_A \cos2\beta )^2}{2p^2(p^2 - m^2_A)(p^2 - m^2_h)(p^2 - m^2_H)}\ . 
\end{align}
In this last expression, the $Z$-boson mass squared, $M^2_Z$, has been
replaced with an effective mass parameter $\widetilde{M}^2_Z$. This
enables us to take into account the quantum corrections to the quartic
coupling $(H^\dagger_u H_u)^2$, in agreement with our simplified
approach discussed in Appendix~\ref{AppendixHiggsSector}.

\section{Loop Correction Coefficients}\label{AppendixCoefficients}

Here we list key auxiliary expressions that we have used in
Section~\ref{sec:1loop}. For convenience, we define
\begin{align}
\frac{1}{v_2}\Big[g(M_{\chi^0}\, \text{adj} \mathcal{F})_{\tilde{h}^0_u \tilde{W}^3}-g'(M_{\chi^0}\, \text{adj} \mathcal{F})_{\tilde{h}^0_u \tilde{B}}\Big] &= A k^6 + B k^4 + C k^2 + D\,, \label{bracket1}
\end{align}
where $\mathcal{F}$ is a $4\times 4$ matrix given in~\eqref{eq:calF} and the
coefficients $A,B,C,D$ are found to be
\begin{subequations}
    \label{coefs}
\begin{align}
    A =& -\frac{1}{2}(g'^2 + g^2),\label{coef1}\\[1ex]
    B =& \frac{1}{2}M_Z^2 + \frac{g'^2}{2}(|\mu|^2 + |M_2|^2 - \mu M_1
         \cot{\beta})+\frac{g^2}{2}(|\mu|^2 + |M_1|^2 - \mu M_2
         \cot{\beta})\label{coef2},\\[1ex] 
    C =& \frac{|\mu|^2}{2}[\mu\cot\beta(g'^2 M_1 + g^2
         M_2)-g'^2|M_2|^2 - g^2 |M_1|^2 - 2v^2_d(g'^2 + g^2)]
         \nonumber\\[2ex] 
    &- (g'^2 M^*_2 + g^2 M^*_1)[(v^2_d + v^2_u)(g^2 M_1 + g'^2 M_2) -
      4\mu M_1 M_2 \cot\beta]\label{coef3},\\[2ex] 
    D =& \frac{|\mu|^2}{4}(g'^2 M^*_2 - g^2 M^*_1)[(g'^2 M_2 + g^2
         M_1)v_d v_u - 2\mu M_1 M_2]\cot{\beta}\,.\label{coef4} 
\end{align}
\end{subequations}

Similarly, we may define
\begin{align}
\frac{1}{2}\Big[g'^2(M_{\chi^0}\, \text{adj} \mathcal{F})_{\tilde{B} \tilde{B}} + g^2(M_{\chi^0}\, \text{adj} \mathcal{F})_{\tilde{W}^3 \tilde{W}^3}\Big] &= A' k^6 + B' k^4 + C' k^2 + D'\,,\label{bracket2}
\end{align}
where
\begin{subequations}\label{coefs'}
\begin{align}
    A' =&\,\frac{1}{2}(g'^2 M_1 + g^2 M_2)\,,\label{coefp1}\\
    B'  =& -\frac{g'^2}{2}M_1 |M_2|^2 - \frac{g^2}{2}M_2 |M_1|^2 -
           |\mu|^2(g'^2 M_1 + g^2 M_2) -\frac{g'^2g^2}{4}(M_1 +
           M_2)(v^2_d + v^2_u)\nonumber\\ 
    &+ \frac{\mu^*}{4}(g'^4 + g^4)v_d v_u\,,\label{coefp2}\\
    C' =&\, \frac{|\mu|^4}{2}(g'^2 M_1 + g^2 M_2) -
          \frac{\mu^*}{4}|\mu|^2 (g'^4 + g^4)v_d v_u -
          \frac{g'^2}{4}|M_2|^2\mu^*(g'^2 v_d v_u - 4\mu M_1)
          \nonumber\\
&- \frac{g^2}{4}|M_1|^2\mu^*(g^2 v_d v_u - 4\mu
                       M_1) + \frac{|\mu|^2}{4}g'^2 g^2(M_1 +
                       M_2)(v^2_d + v^2_u) \nonumber\\
&-
          \frac{\mu^*}{4}g'^2
          g^2(M_1 M^*_2
          + M^*_1
          M_2)v_d v_u +
          \frac{g'^2g^2}{16}(g^2
          M_1 + g'^2
          M_2)(v^2_d +
          v^2_u)^2
          \nn\\&-
                 \frac{\mu}{2}g^2
                 g'^2
                 M_1 M_2
                 v_d
                 v_u\,,\label{coefp3}\\ 
  D' =& -\frac{|\mu|^2}{4}[\mu^*(g'^2 M_2 + g^2 M_1) - g'^2 g^2 v_d
        v_u][(g'^2 M_2 + g^2 M_1)v_d v_u - 2\mu M_1 M_2]\, .\label{coefp4}
\end{align}
\end{subequations}
Finally, we have used the expression
\begin{align}
\frac{2}{v^2_u}\Big[g'^2(M_{\chi^0}\, \text{adj} \mathcal{F})_{\tilde{h}_u
  \tilde{h}_u} \Big] &= A'' k^6 + B'' k^4 + C'' k^2 +
                       D''\,,\label{bracket3} 
\end{align}
where
\begin{subequations}\label{coefs''}
\begin{align}
    A'' =&\,0\,,\label{coefpp1}\\[2ex]
    B'' =&\, 4(g'^2 M_1^* + g^2 M^*_2) + 8\mu\cot\beta(g'^2 + g^2)\,,\label{coefpp2}\\[2ex]
    C'' =&\,2 g_1^4 \mu  |M_2|^2 v_d^2 \cot\beta + 2\mu g_1^2 M_1 M_2^* \cot^2\beta  \left(g_2^2 v_d v_u - 2 \mu  M_2\right)\nn\\[2ex]
    &+ 2 g_2^4 \mu  |M_1|^2 v_d^2 \cot\beta + 2\mu g_2^2 M_2 M_1^* \cot^2\beta  \left(g_1^2 v_d v_u - 2 \mu  M_1\right), \label{coefpp3}\\[2ex]
    D'' =&\,-8 \mu v_d v_u (g'^2|M_2|^2 + g^2 |M_1|^2)
    -4 v^2_u M^*_1 M^*_2 (g'^2 M_2 + g^2 M_1)\nn\\[2ex]
    &+4 \mu^2 v_d^2 (g'^2 M_1 + g^2 M_2) - 2 (g'^2 + g^2)^2 \mu  v_d^3
      v_u\,. \label{coefpp4}
\end{align}
\end{subequations}


\bibliographystyle{unsrt}
\bibliography{bibs-refs.bib}

\end{document}